\documentclass[preprint,11pt]{JHEP3} 

\JHEPspecialurl{http://jhep.sissa.it/JOURNAL/JHEP3.tar.gz}
\JHEPspecialurl{http://jhep.sissa.it/JOURNAL/JHEP3.tar.gz}
\usepackage{epsfig,multicol,amsmath,slashed}

\def\beq{\begin{equation}}
\def\eeq{\end{equation}}
\def\bea{\begin{eqnarray}}
\def\eea{\end{eqnarray}}

\def\eq#1{{Eq.~(\ref{#1})}}
\def\fig#1{{Fig.~\ref{#1}}}
\newcommand{\bas}{\bar{\alpha}_S}
\newcommand{\as}{\alpha_S}

\newcommand{\mn}{{\mu\nu}}
\newcommand{\NA}{\Lb\,\frac{2\pi\bas^2}{N_c}\Rb}

\newcommand{\tr}{{\rm tr}}

\newcommand{\Lb}{\left(}
\newcommand{\Rb}{\right)}
\newcommand{\s}{\slashed}
\newcommand{\p}{I\!\!P}

\newcommand{\ikk}{\int\!\frac{d^4k}{\Lb 2\pi\Rb^4}}

\parskip=\itemsep               

\newcommand{\h}{\frac{1}{2}}

\newcommand{\kv}{\vec{k}}
\newcommand{\qv}{\vec{q}}

\renewcommand{\theequation}{\thesection.\arabic{equation}}

\newcommand{\Itn}{\int^{\infty}_{-\infty}\!d\nu}
\newcommand{\Itnpr}{\int^{\infty}_{-\infty}\!d\nu'}

\newcommand{\Itnp}{\int^{\infty}_{-\infty}\!\frac{d\nu}{2\pi\,i}}

\newcommand{\pItn}{\Itn \, \nu^2\,\,
\,\prod^2_{i=1}\Itn_i\nu^2_i\,\lambda\Lb\,\nu_i\Rb\,\,
\,\Itnpr_i\nu'^2_i\,\lambda(\nu'_i)\,}
\newcommand{\pItnprr}{\int^{\infty}_{-\infty} d\nu\, \nu^2\,\,
\,\prod^2_{i=1}\,\int^{\infty}_{-\infty} d \nu_i
\nu^2_i\,\lambda\Lb\,\nu_i\Rb\,\,\int^{\infty}_{-\infty} d \nu'_i
\,\nu'^2_i\,\lambda(\nu'_i)\, \, \,\int^{\infty}_{-\infty} d
\nu'\,\nu'^2\,}
\newcommand{\GVc}{\mbox{GeV}/\mbox{c}^2}
\newcommand{\mb}{\mbox{mb}}
\newcommand{\intf}{\int^{\infty}_{-\infty}}
\newcommand{\sigmaf}{2\pi^9\NA^2\Gamma_P\Lb k,q-0\Rb\Gamma_P\Lb
k_0,q=0\Rb\sigma\Lb M_W\Rb\sigma\Lb M_{b\bar{b}}\Rb}
\title{\LARGE \bf Two parton shower  background for associate W Higgs production }
\author{\large  E. ~Levin\thanks{Email: leving@post.tau.ac.il,
levin@mail.desy.de;} \,\,\,\,and \,\,J.~Miller\thanks{Email:
jeremymiller@london.com,jeremymi@post.tau.ac.il;}\,\, \\
Department of Particle Physics, School of Physics and Astronomy\\
Raymond and Beverly Sackler
 Faculty
of Exact Science\\  Tel Aviv University, Tel Aviv, 69978, Israel}

\abstract{ The estimates of  the background for the  associate W
Higgs production, which stems from the two parton shower production.
It is about $1 \div 2.5$ times larger than the signal. However, this
background does not depend on the rapidity difference between the W
and the $b \bar{b}$ pair,  while the signal peaks when the rapidity
difference is zero. The detailed calculations for the enhanced
diagrams' contribution to this process, are presented, and it is
shown that the overlapping singularities, being important
theoretically, lead to a negligible contribution for the LHC range
of energies.}

 \keywords{Higgs production, BFKL Pomeron, Pomeron loops, correlation function, overlapping singularities, W  inclusive  production, W and quark -antiquark  associate production}

\preprint{  TAUP -2870-08\\
hep-ph/???????\\
\today}
\begin{document}

\numberwithin{equation}{section}

\section{Introduction}
The most important  discovery, that everybody expects at the LHC, is
the discovery of the Higgs boson.  The exclusive process which has
the best experimental signature,  as far as we know, is diffractive
Higgs production.  Having two large rapidity gaps between the Higgs
and the recoiled protons, this process has a minimal background from
QCD  processes without the Higgs. However, the total cross section
for diffractive Higgs production turns out to be very small, namely
about 3 fb (see detailed estimates by the Durham group, in Ref.
\cite{DG}).  The largest cross section out of all the different
processes for Higgs production, is the cross section for inclusive
Higgs production, which reaches 40 - 50 pb (assuming that the Higgs
mass $M_H \approx 100\,\GVc$) .

\FIGURE[ht]{\begin{minipage}{80mm}
\centerline{\epsfig{file=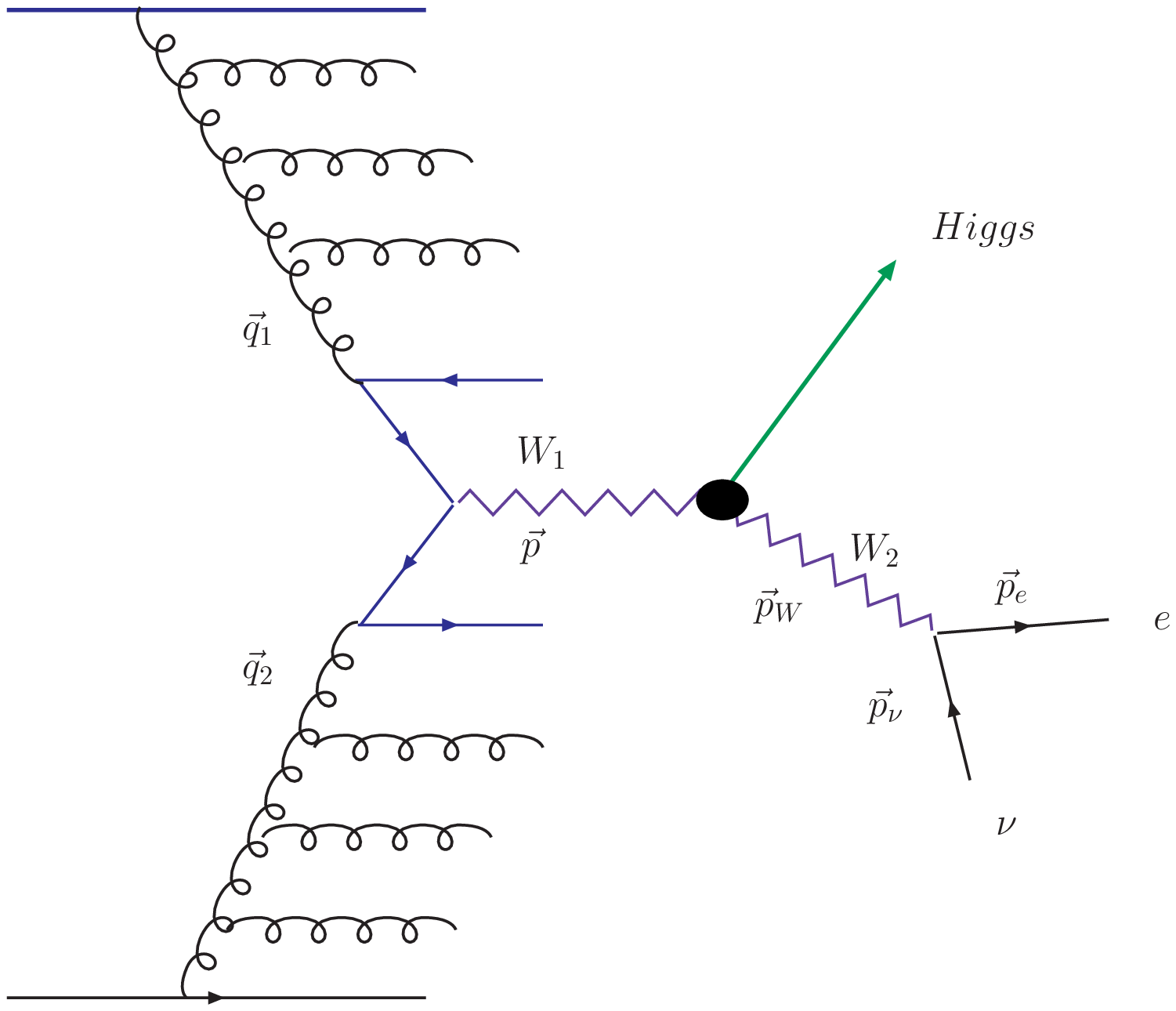,width=75mm}}
\end{minipage}
\caption{ The  associate W boson and  Higgs production. }
\label{hwprod} }

Unfortunately, the background for this process turns out to be so
large, that any experimental measurement of the Higgs in an
inclusive process, looks rather problematic. In his talk at GGI WS
(Florence 2007),  Hannes Jung made the suggestion of measuring the
associate Higgs production, which is the inclusive production of the
Higgs together with the W boson. This process has a cross section
which is about 1.5\%  of that of the cross section for inclusive
Higgs production, but this smallness could be compensated by its
improved experimental signature. This process yields the production
of the W boson and the Higgs, at more or less the same rapidity
values (see \fig{hwprod}). The value for the cross section for this
process, can be found only to within an accuracy of 30\%, due to
large uncertainties in the values of the structure functions.
However, the ratio

\beq \label{WH2W} R_1\,\,\,= \,\,\,\,\frac{\sigma\left( pp \,\,\to
\,\,WH \,\,+ \,\,X \right)}{\sigma\left( pp\,\, \to \,\,W\,\, +
\,\,X\right)}\,\,\approx\,\,3.5\,\times\,10^{-8} \eeq

is known to within 7\% accuracy. In this paper, we estimate the
cross section of the main background process, namely, the associate
production of the W boson and the $ b \bar{b}$ pair, which has a
mass close to $M_H$, and is produced at more or less the same value
of rapidity. It is easy to see, that if we assume that the W boson
and the $b \bar{b} $ pair are produced independently of the
different parton showers, the ratio of the signal to the background
can be estimated using the following equation

 \bea \label{RSB}
&&R\,\,\,=\,\,\,\frac{\sigma\left( pp \,\,\to \,\,W H \,\,+ \,\,X
\right)}{\sigma\left( pp\,\, \to \,\,W\,\, + \,\,[b \bar{b}] \,\,\,+
X\right)}\,\,\approx\,\,\frac{R_1}{\frac{\sigma( pp \to  [b \bar{b}]
\,\,+\,\,X)}{\sigma_{tot}}} \nonumber \eea

This simple estimate for the cross section for $b \bar{b}$
production, gives

 \beq \label{EXSBB} \sigma( pp \to  [b \bar{b}]
\,\,+\,\,X)\,\,\propto\,\,\dfrac{\as^2(M_H)}{M^2_H} \eeq

 and, using
$ \sigma_{tot} = 110\,\mbox{mb}$ (see Ref. \cite{TXS}), we obtain

\beq \label{II1} R\,\,\sim\,\,9 \div 10 \eeq

 However, we see two
sources for an increase of the $b \bar{b}$ production cross section.
These are the energy growth of the cross section due to the gluon
structure function, and the positive correlation between the W boson
and the $b \bar{b}$ production, which has been seen at the Tevatron
\cite{DPXS}.   The estimates of
 both sources, are the subject of this paper.\\

 This paper  is organized as follows.
In the next section, we calculate the cross sections for inclusive W
boson and  $b \bar{b}$ production,  with a particular focus on
    the energy dependence of the cross section.

 The main section of this paper is the third, where we consider the correlation function for inclusive  W boson
   and $b \bar{b}$  production. We show that this correlation function can be reduced to
   the calculation of the specific enhanced diagrams, for the BFKL Pomeron interaction.  The detailed  calculations of the enhanced diagrams are presented in this section, which include the integration over the entire kinematic region.  In particular, the overlapping singularities are taken into account.
     Calculating this diagram, and comparing this calculation with the experimental data obtained at the Tevatron,
     we will be able to provide a reliable estimate in section 4   of the ratio of  \eq{RSB}, at
     the LHC range of energies. In the conclusion we summarise our results.

\section{One parton shower contribution: one BFKL Pomeron exchange and inclusive cross section}

In this section, we will discuss the calculation of the inclusive
cross section. Due to the factorisation theorem \cite{FT,KTF}, and
the AGK cutting rules \cite{AGK}, only the production from one
parton shower contributes to the single inclusive cross section. We
begin with a discussion of the one BFKL Pomeron exchange amplitude.

\FIGURE[ht]{\begin{minipage}{110mm}
\centerline{\epsfig{file=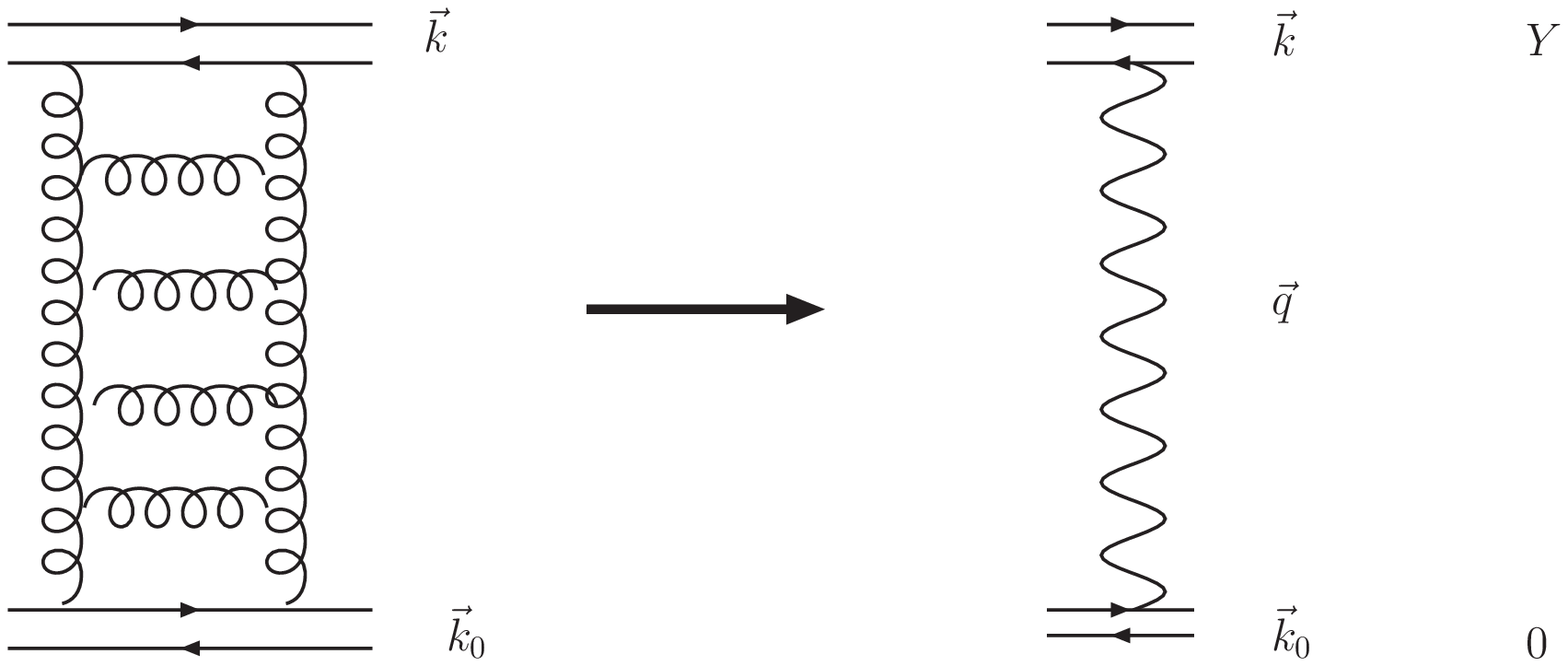,width=95mm}}
\end{minipage}
\caption{ The  exchange of the single BFKL Pomeron. } \label{1p} }

In the calculations that follow, let $\vec{k}_{i}$ denote the
momenta conjugate to the size of the interacting dipoles.
$\vec{q}_i$ are the momenta transferred along the Pomerons. The
diagram of \fig{1p} shows the scattering of the dipole with rapidity
$Y$, and transverse momentum $\vec{k}$, off the dipole at rapidity
$Y=0$, which carries transverse momentum $\vec{k}_0$, due to the
exchange of one Pomeron. In this notation, the amplitude of \fig{1p}
has the expression\cite{BFKL}

 \bea \label{1P1g}
A\Lb 1\p\Rb\,\,&=&\,\,\sum^{+\infty}_{n= - \infty}\,\oint_C\,d\gamma
\,g\Lb \vec{k},\vec{q};\gamma_n \Rb \,e^{\omega(\gamma_n) (Y -
Y')}\,h\Lb\,\gamma_n\Rb\,\,\,
g\Lb\vec{k}_0, \vec{q};\tilde{\gamma}  \Rb\notag\\
\mbox{where}\,\,\,\,\,\,\,\,\,h\Lb\,\gamma_n\Rb\,&=&\,\Lb\,\gamma_n\,-\,\frac{1}{2}\Rb^2\lambda\Lb\,\gamma_n\Rb\,\,\,\,\,\,\,\mbox{and}\,\,\,\,\,\,\,\,\,\,\,\,\,\,\,\lambda\Lb\,\gamma_n\Rb\,=\,\Lb\,\gamma_n\Lb\,\,1-\gamma_n\Rb\,\Rb^{-2}
\eea

where the conformal variable $\gamma_n$, is defined in \eq{gamdef}
as

\bea
\gamma_n\,=\,\frac{1+n}{2}\,+\,i\nu\,\,\,\,\,\,\,\,\,\,\,\,\,\,\,\,\,\,\,\,\,\,\,\,\,\,\,\,\,\,\,\,\,\,\,\,\,\,\,\,\,\,\,\tilde{\gamma}_n\,=\,1\,-\,\gamma_n^{\ast}\,=\,\frac{1-n}{2}+i\nu\label{gamdef}
\eea

\FIGURE[ht]{\begin{minipage}{80mm}
\centerline{\epsfig{file=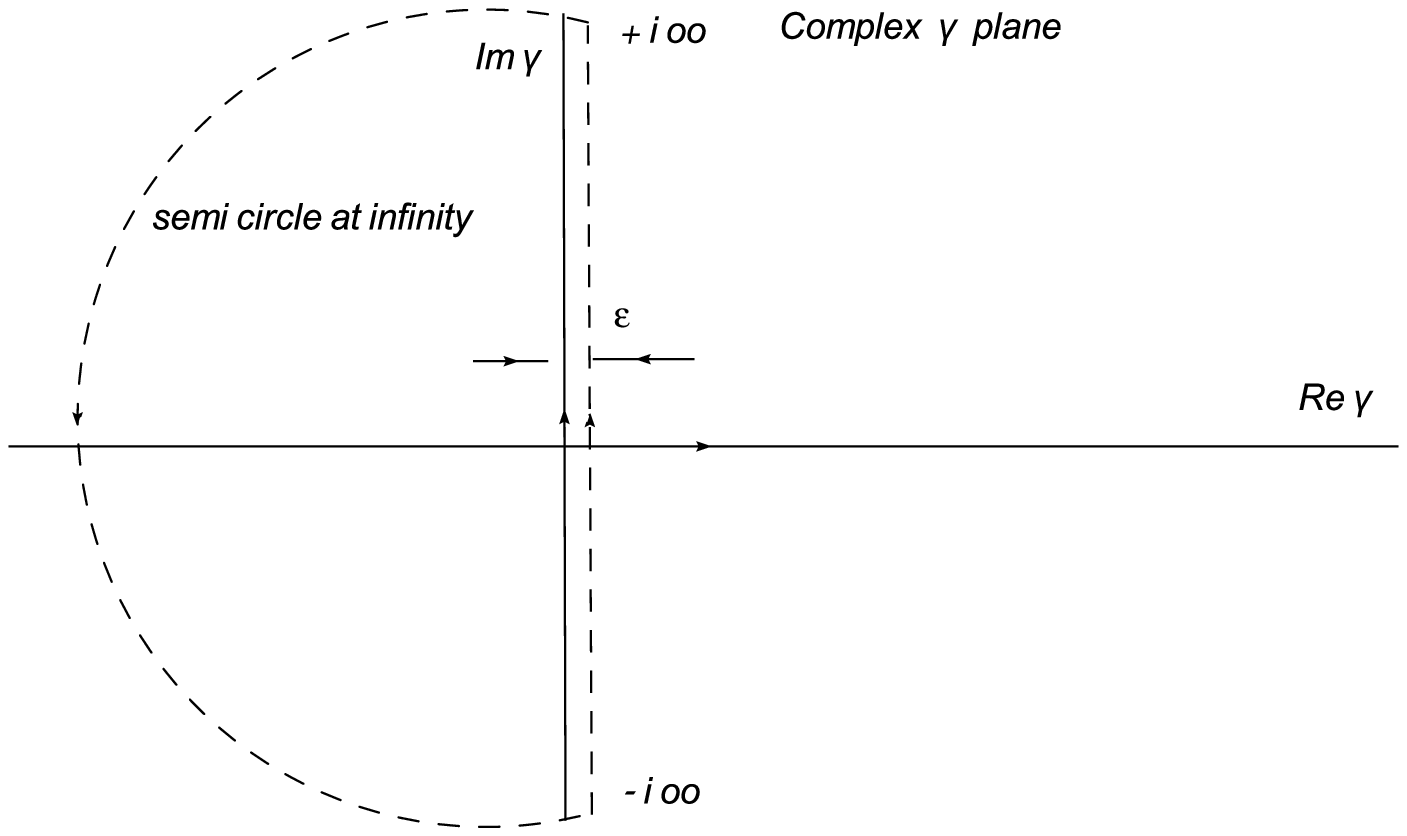,width=75mm}}
\end{minipage}
\caption{Contour enclosing singularities for integration over the
conformal variable $\gamma$.  } \label{graph} } The contour $C$
consists of the imaginary $\gamma$ axis from $\pm\,i\infty$, and the
semi circle at $\infty$, to the left of the imaginary $\gamma$ axis,
which encloses all singularities in the
integrand of \eq{1P1g} (see \fig{graph}). Since\\
 $g\Lb
\vec{k},\vec{q};\gamma  \Rb\,g\Lb\vec{k}_0, \vec{q};\tilde{\gamma}
\Rb\,\,\propto\,\,\Lb k^2/k^2_0 \Rb^\gamma$ for $k^2 > k^2_0$, the
integrand vanishes on the semi circle at $\pm\,i\infty$, such that
it is sufficient just to replace

 \beq
\oint_C\,d\gamma_n\,\rightarrow\,\int^{\epsilon\,+i\infty}_{\epsilon
- i\,\infty}d\gamma_n\,\label{ig} \eeq

In evaluating the integral of \eq{1P1g}, and in particular for the
first enhanced diagram, the calculations that follow will be more
economical if the integrals are expressed in terms of the variable
$\nu$, instead of the variable $\gamma_n$, (where the relationship
between $\gamma_n$ and $\nu$ is given in \eq{gamdef}). For the
integration limits
$\epsilon+i\infty\leq\gamma\leq\epsilon+i\,\infty$ (as
$\epsilon\to\,0$), then \eq{gamdef} gives the corresponding limits
of integration for the variable $\nu$ as $-\infty\leq\nu\leq\infty$,
and one should sum over all real positive integers $n$ in
\eq{gamdef}. In this way in \eq{1P1g}, one can replace the
integration over $\gamma_n$ with the integration over $\nu$ using
the notation

\beq \sum^{\infty}_{n=-\infty}\,\int^{\epsilon\,+i\infty}_{\epsilon-
i\,\infty}d\gamma_n\,\,=\,\sum^{\infty}_{n=-\infty}\,\Itn\label{gamnu}
\eeq

The notation $\sum^{\infty}_{n=-\infty}\Itn$, corresponds to the
integration over the quantum numbers associated with the continuous
unitary variable irreducible representations of
$SL\,\Lb\,2\,,\,C\Rb\,$ , defined in \eq{gamdef}. The energy levels
are the  $SL\,\Lb\,2\,,C\Rb\,$  eigenvalues of the BFKL kernel,
given by

 \bea
\omega\Lb\,\gamma_n\Rb\,\,\,\,\,\,\,=\,\,\,\,\,\,\,\,\,\,\,\bas\left\{\psi\Lb\,1\Rb\,-\Re\,\Lb\,\psi\Lb\,\gamma_n\Rb\,\Rb\,\right\}\,\,\,\,\,&=&\,\,\,\,\,\,\,\bas\left\{2\psi\Lb\,1\Rb\,-\psi\Lb\,\gamma_n\Rb\,-\psi\Lb\,1-\gamma_n\Rb\,\right\}\label{om}\\
\notag\\
&=&\,\,\,\,\,\,\,\,\bar{\alpha}_S \left\{ \psi(1) - Re{\,
\psi\left(\frac{|n| + 1}{2} + i \nu\right)} \right\}\label{OM} \eea

where $\psi\Lb f\Rb\,\,=\,d\,/\,df\,\Lb\,\ln\Gamma\Lb\,f\Rb\,\Rb\,$,
and $g\Lb \vec{k}, \vec{q};\gamma_n  \Rb $, are the eigenfunctions.
Thus, the one BFKL pomeron exchange amplitude shown in \fig{1p}, has
this expression in $\nu$ notation

 \bea \label{1P1}
A\Lb 1\p\Rb\,\,&=&\,\,\Itn\,g_P\Lb \vec{k}, \vec{q};n=0,\nu  \Rb \,e^{\omega(n=0,\nu) (Y - Y')}\,\nu^2\,\lambda(n=0,\nu)\,\,g_P\Lb \vec{k}_0, \vec{q};n=0,-\nu  \Rb \nonumber\\
&=&\Itn \,g_P\Lb \vec{k}, \vec{q}; \nu  \Rb \,e^{\omega(\nu) (Y -
Y')}\,\nu^2\,\lambda\Lb 0,\nu\Rb
 \,g_P\Lb \vec{k}_0, \vec{q};-\nu  \Rb
\eea

 where in terms of $\nu$

\begin{equation} \label{BFKLLA}
\lambda(n,\nu)\,=\frac{1}{[ \frac{( n + 1)^2}{4} +  \nu^2] [\frac{(n
- 1)^2}{4} + \nu^2]}
\end{equation}

In \eq{1P1}, it is assumed that $n=0$, in order to find the largest
contribution at high energy. From now onwards, the notation
$\omega(\nu) = \omega(n=0,\nu)$ will be used (see \eq{OM}), and
$\lambda(\nu) = \lambda(n=0,\nu)$ and $g\Lb \vec{k}, \vec{q}; \nu
\Rb = g\Lb \vec{k}, \vec{q};n=0, \nu \Rb$. In \eq{1P1}, $
g_P\Lb\,q,k;n=0,\nu\Rb\,\,=\,\,\Gamma_P(k,q)\,\,g\Lb\,q,k;n=0,\nu\Rb\,$
and $g\Lb\,q,k;n=0,\nu\Rb\,$   has been calculated in
Ref.\cite{LMP}, which takes the form the form

\bea \,
g\Lb k,q;\nu\Rb\,\,&=&\frac{\Lb q\bar{q}\Rb^{i\nu}}{\vert\,k\vert\,}\left\{C_1\Lb \nu\Rb\Lb\frac{q^2}{16\,k^2}\Rb^{-i\nu}F\Lb \nu\Rb-C_2\Lb \nu\Rb\Lb\frac{q^2}{16\,k^2}\Rb^{i\nu}F\Lb -\nu\Rb\right\}\label{G}\\
\mbox{where}\,\,\,\,\,\,\,\,F\Lb \nu\Rb\,&=&\,_2F_1\Lb\,\frac{1}{4}-\h i\nu,\frac{3}{4}-\h\,i\nu,1-i\nu,\frac{q^2}{4k^2}\Rb_2F_1\Lb\frac{1}{4}-\h i\nu,\frac{3}{4}-\h\,i\nu,1-i\nu,\frac{\Lb q^{\ast}\Rb^2}{4\Lb k^{\ast}\Rb ^2}\Rb\notag\\
\mbox{ and where}\,\,\,\,\,\,C_1\Lb
\nu\Rb&=&\,\frac{2^{-2i\nu}\pi^2}{-i\nu}\frac{\Gamma^2\Lb\h-i\nu\Rb\Gamma\Lb
i\nu\Rb}{\Gamma^2\Lb \h+i\nu\Rb\Gamma\Lb
-i\nu\Rb}\,\,\,\,\,\,\,C_2\Lb
\nu\Rb\,=\,\frac{2^{-2i\nu}\pi^2}{-i\nu}\frac{\Gamma^2\Lb1-i\nu\Rb\Gamma\Lb
i\nu\Rb}{\Gamma^2\Lb 1+i\nu\Rb\Gamma\Lb -i\nu\Rb}\label{GF}\eea

>From this definition, one notes the following interesting property
of $C_1\Lb \nu\Rb$, namely

 \beq
C_1\Lb \nu\Rb\,C_1\Lb -\nu\Rb=\,\frac{\pi^4}{\nu^2}\label{CC}
 \eeq

 In the limit that $q\,\ll\,k$, then the function $F\Lb\,\nu\Rb\,$
defined in \eq{GF} tends to unity, and \eq{G} reduces to

\beq \label{Ge0}
g\Lb\,k\,,\,q\,,\,n=0\,,\,\nu\Rb\,=\,\frac{C_1\Lb\,\nu\Rb\,}{\vert\,k\vert\,}\Lb\,16k^2\Rb^{i\nu}
\eeq

$\Gamma_P(k,q)$ is the so called impact factor, which is equal to

\beq \label{IMFA} \Gamma_P(k,q)\,\,\,=\,\,\sum_i\,\int\,\prod_i\,d
^2 x_i\,d^2 y_i\, |\Psi \Lb \{x_i,y_i\} \Rb|^2\, e^{ \vec{q} \cdot
\h (\vec{x}_i\,-\,\vec{y}_i)}\,\,\Lb 1\,-\, e^{ \vec{k} \cdot
(\vec{x}_i\,-\,\vec{y}_i)}\Rb \eeq

 where $\Psi\Lb \{x_i,y_i\} \Rb$
is the wave function of the proton written in terms of the
colourless dipoles, where $x_i$ and $y_i$ are the coordinates of the
dipole $i$ (here $|\vec{x}_i\,-\,\vec{y}_i|$ is the dipole size). It
is convenient to calculate the amplitude for single pomeron exchange
of \eq{1P1}, for the case of $\vec{q}\,=\,0$, since we are
considering here the contribution of the BFKL Pomeron to the total
cross section. Hence, assuming that $k^2_0 \,\ll \,k^2$, and using
the expression of \eq{G} for $g\Lb\,q,k;n=0,\nu\Rb\,$

 \bea \label{1P2a} A\Lb 1\p
\Rb\,\,&=&\,\, \,\frac{\pi^4}{\Lb k^2\,k^2_0\Rb^{1/2}}\,\,
\Gamma_P\Lb k,q=0 \Rb\,\Gamma_P\Lb k_0,q=0 \Rb\,\,
P\Lb k;k_0|Y-Y'\Rb \notag\\
\mbox{where}\,\,\,\,\,\,\,\,\,\,\,P\Lb k;k_0|Y-Y'\Rb \,&=&\,\Itnp\,
\,e^{\omega(\nu) (Y - Y')}\,\lambda\Lb
0,\nu\Rb\Lb\frac{k^2}{k^2_0}\Rb^{i \nu}\,\,\,\,\,\, \eea

 It should be noted that $P\Lb k;k_0|Y-Y'\Rb$
is the dimensionless amplitude for the single BFKL pomeron exchange,
where as $A\Lb\,1\p\Rb\,$ is related to the cross section, and it
has the dimensions of $\mbox{GeV}^{-2}$. The integration over $\nu$
can be evaluated by expanding the BFKL function
$\omega\Lb\,\nu\Rb\,$, around the saddle point $\nu=0$ (see \eq{OM})

\beq  \label{OMe} \omega\Lb\,\nu\Rb\,
=\omega\Lb\,0\Rb\,\,\,-\,\,\frac{1}{2}\nu^2\omega"\Lb\,0\Rb\,\eeq

Then the integration over $\nu$ in \eq{1P2a}, reduces to a Gaussian
integral, and evaluating the $\nu$ integration, gives the expression
for single Pomeron exchange as the expression

\FIGURE[ht]{ \centerline{\epsfig{file=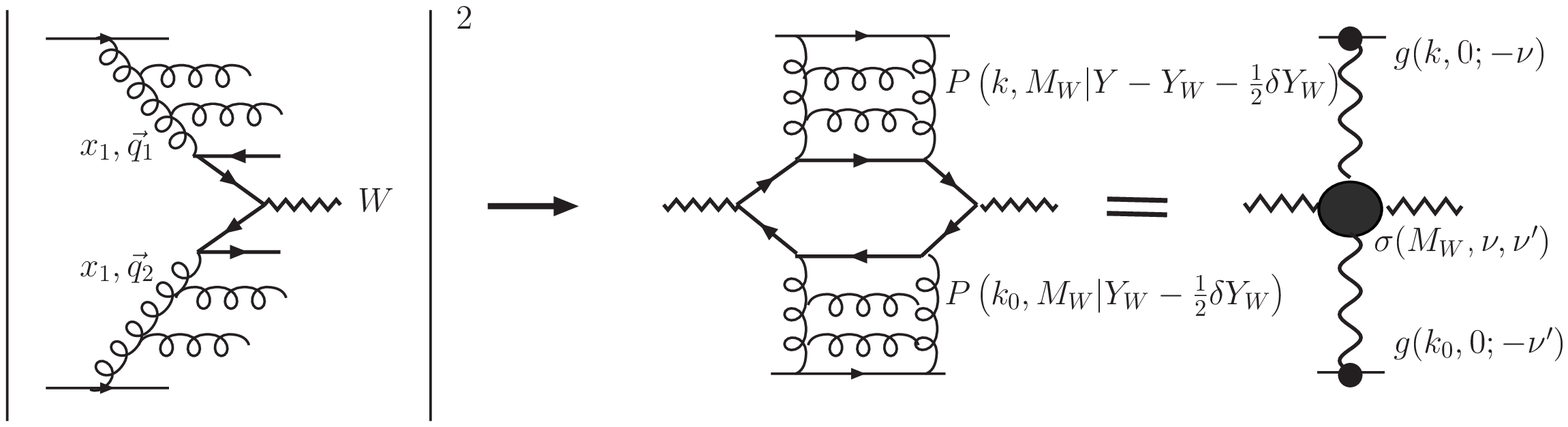,width=140mm}}
\caption{ The  inclusive production  in the  BFKL Pomeron approach.
} \label{incw} }

\bea [A\Lb\,1\p\,\Rb\,]_{\nu\to\,0}&=&\,\frac{\pi^4\,\Gamma_P\Lb
k,q=0 \Rb\,\Gamma_P\Lb k_0,q=0
\Rb\,\,}{\Lb\,k_0^2k^2\Rb^{\frac{1}{2}}}P\Lb
k;k_0\vert\,Y-Y'\Rb\label{1P2}\\
\mbox{where}\,\,\,\,\,\,\,\,\,\,\,\,P\Lb
k;k_0\vert\,Y-Y'\Rb\,&=&\,e^{\omega\Lb\,0\Rb\,Y}\Lb\,\frac{2\pi}{\omega"\Lb\,0\Rb\,Y}\Rb^{\frac{1}{2}}\label{1P2'}\eea

It should be stressed, that the value of $\nu$ in this integral is
small, namely $\Lb \nu\,= \,\ln (k^2/k^2_0)/(\omega"(0) Y) \ll
1\Rb$, justifying  our expansion in \eq{OMe}. It should be noted,
that in \eq{1P2a},   contains as the initial condition at $Y - Y'
=0$, the exchange of two gluons in the Born Approximation of
perturbative QCD. Indeed,  in this case, the contour in $\nu$  can
be closed on the singularities of $\lambda (n\,=\,0,\nu)$, which are
the poles of the second order at $ \nu = \pm i \h$. Considering the
case when $k^2 > k^2_0$, we close the contour on the point $\nu = +
\h\,i$, and obtain the following expression

 \beq \label{ICM} P\Lb k;k_0|Y-Y' \to 0
\Rb\,\,\,\longrightarrow\,\,\,\sqrt{\frac{k^2_0}{k^2}}\,\,\ln \Lb
k^2/k^2_0 \Rb \eeq which leads to \beq \label{ICM1}
A\Lb\,1\p\,\Rb\,\,=\,\,\frac{\pi^4\,\Gamma_P\Lb k,q=0
\Rb\,\Gamma_P\Lb k_0,q=0 \Rb\,\,}{k^2}\,\,\,\ln \Lb k^2/k^2_0 \Rb
\eeq

 \eq{ICM1} is the answer for the Born approximation of
perturbative QCD, (namely, for the case of two gluon exchange). For
the case of calculating the inclusive cross section, we can use the
$k_t$ factorisation theorem \cite{FT,KTF}, using the Mueller
technique \cite{MUIN}, which allows one to reduce the calculation of
the inclusive cross section to the case of single Pomeron exchange,
using the optical theorem
 (see \fig{incw}, where the inclusive
production of the W boson is shown in the Mueller technique
\cite{MUIN}). Indeed, according to the factorisation theorem, the
inclusive production can be written as \cite{INXS}

 \beq \label{INC1}
\frac{d \sigma}{d Y_W}\,\,=\,\,\int d x_1 d x_2\,\int d^2 q_1 \,d^2
q_2\,\phi(x_1,q_1)\,\phi(x_2,q_2)\,\sigma(M_W,q_1,q_2)\,\delta( x_1
x_2 s - M^2_W) \eeq

 where $s$ is the energy of the collision, and
$\phi(x_i,q_i)$ is the un-integrated structure function, which is
related to the gluon structure function in the following way

\beq \label{XGPHI} x_iG(x_i,q^2)\,\,=\,\,\int^{q^2}\,d k^2
\phi(x_i,k^2) \eeq

 where the $x_i$ are equal to

 \beq \label{XI}
x_1\,\,\,=\,\,\frac{M_W}{\sqrt{s}}\,e^{
Y_W}\,\,;\,\,\,\,\,\,\,\,\,\,\,\,\,\,\,\,\,x_2\,\,\,=\,\,\frac{M_W}{\sqrt{s}}\,e^{
-Y_W}\,\,; \eeq

 where $Y_W$ is the rapidity of the W boson in the
center of mass frame (c.m.f.). The un-integrated structure function
at high energy can be rewritten through the Pomeron exchange, while
the hard cross section in \eq{INC1} is calculated in the appendix.
The integrals over $q_1$ and $q_2$ in \eq{INC1},  are convergent  at
$q_1 \approx M_W$ and $q_2 \approx M_W$, and because of this, one
can rewrite \eq{INC1} in a simpler form, namely

\bea &&\frac{d \sigma_{incl}\Lb \ln\Lb s/M^2_{b , \bar{b}}\Rb ;Y_{b
\bar{b}}; 0 \Rb}{d y_{b \bar{b}}}\,\,=\label{FT1} \\
&& \sigma\Lb  M_{b \bar{b}}\Rb\,\,\,\Gamma_P\Lb k,q=0
\Rb\,\Gamma_P\Lb k_0,q=0 \Rb\,\,\sqrt{\frac{M^4_{b
\bar{b}}}{k^2\,k^2_0}}\,\,
P\Lb k,M_{b , \bar{b}}| Y - Y_{b \bar{b}} - \h \delta Y_{b \bar{b}}\Rb \,\,\,P\Lb k_0,M_{b , \bar{b}}|  Y_{b \bar{b}} - \h \delta Y_{b \bar{b}}\Rb \notag \\
&&\frac{d \sigma_{incl}\Lb \ln\Lb s/M^2_{W}\Rb ;Y_{W}; 0 \Rb}{d
Y_{W}}\,\,=\label{FT}\\
&&\sigma\Lb  M_{W}\Rb\,\,\,\,\,\Gamma_P\Lb k,q=0 \Rb\,\Gamma_P\Lb
k_0,q=0 \Rb\,\,\sqrt{\frac{M^4_{W}}{k^2\,k^2_0}}\,\,P\Lb k,M_{W}| Y
- Y_{W} - \h \delta Y_{W}\Rb \,P\Lb k_0,M_{W}|  Y_{W} - \h \delta
Y_{W}\Rb  \notag \eea

 In \eq{FT}, $\sigma\Lb\,M_W\Rb$ is the squared amplitude of the quark hexagon
  contribution shown in \fig{hex}, and $\sigma\Lb\,M_{b\bar{b}}\Rb$  is the
   squared amplitude of the quark subprocess contribution shown in \fig{bbin}.
   Both have the dimensions of cross section, namely
   $\sigma\Lb\,M_W\Rb$ and $\sigma\Lb\,M_{b\bar{b}}\Rb$ have
   dimensions of $\mbox{GeV}^{-2}$
   (see \eq{WPR} and \eq{BBPR} respectively). $Y = \ln (s/m^2),$ and $Y_W$ and $Y_{b
\bar{b}}$ are respectively the rapidity values in the laboratory
frame of the W boson, and the center of mass of the quark anti-quark
pair. $\delta Y_W\equiv \ln(M^2_W/m^2)$ and $\delta Y_{b \bar{b}}
\,\equiv \,\ln(M^2_{b \bar{b}}/m^2)$ are the rapidity windows
occupied by the W boson and the $[b\bar{b}]$ pair, respectively. The
arguments in the single Pomeron amplitudes in \eq{FT}, follow
directly from \eq{XI}. For $P$, we can use \eq{1P2a}, for example

\beq \label{INC2} P\Lb k,M_{W}| Y - Y_{W} - \h \delta
Y_{W}\Rb\,\,\,\,=\,\,\,\,\,\,\Itnp \,e^{ \omega(\nu)\,\Lb Y - Y_W -
\h \delta Y_W\Rb}\,\,\Lb \frac{M^2_W}{k^2} \Rb^{i \nu} \eeq

Using \eq{ICM},  one can see that at low energies, $Y \,\to\, \delta
Y_W$, and \eq{FT} reduces to the expression

 \beq \label{BAR}
  \frac{d \sigma_{incl}\Lb \ln\Lb s/M^2_{W}\Rb ;Y_{W}; 0 \Rb}{d
Y_{W}}\,\longrightarrow\,\, \,\,\,\,\,\Gamma_P\Lb k,q=0
\Rb\,\Gamma_P\Lb k_0,q=0 \Rb\,\ln\Lb M^2_W/k^2 \Rb\,\ln\Lb
M^2_W/k^2_0\Rb\,\,\sigma\Lb  M_{W}\Rb \eeq

 which is exactly the same
as the expression for the cross section in the Born approximation of
perturbative QCD, ( see \eq{WPRI}),   with $k^2 = k^2_0 = q^2_{\bot,
\mbox{min}}$, and  $\Gamma_P\Lb k,q=0\Rb \,=\,\,\bas (k^2)$.  It is
easy to see, that \eq{FT1}  has the form of \eq{BBPRI}, in the low
energy limit.

\DOUBLEFIGURE[t]{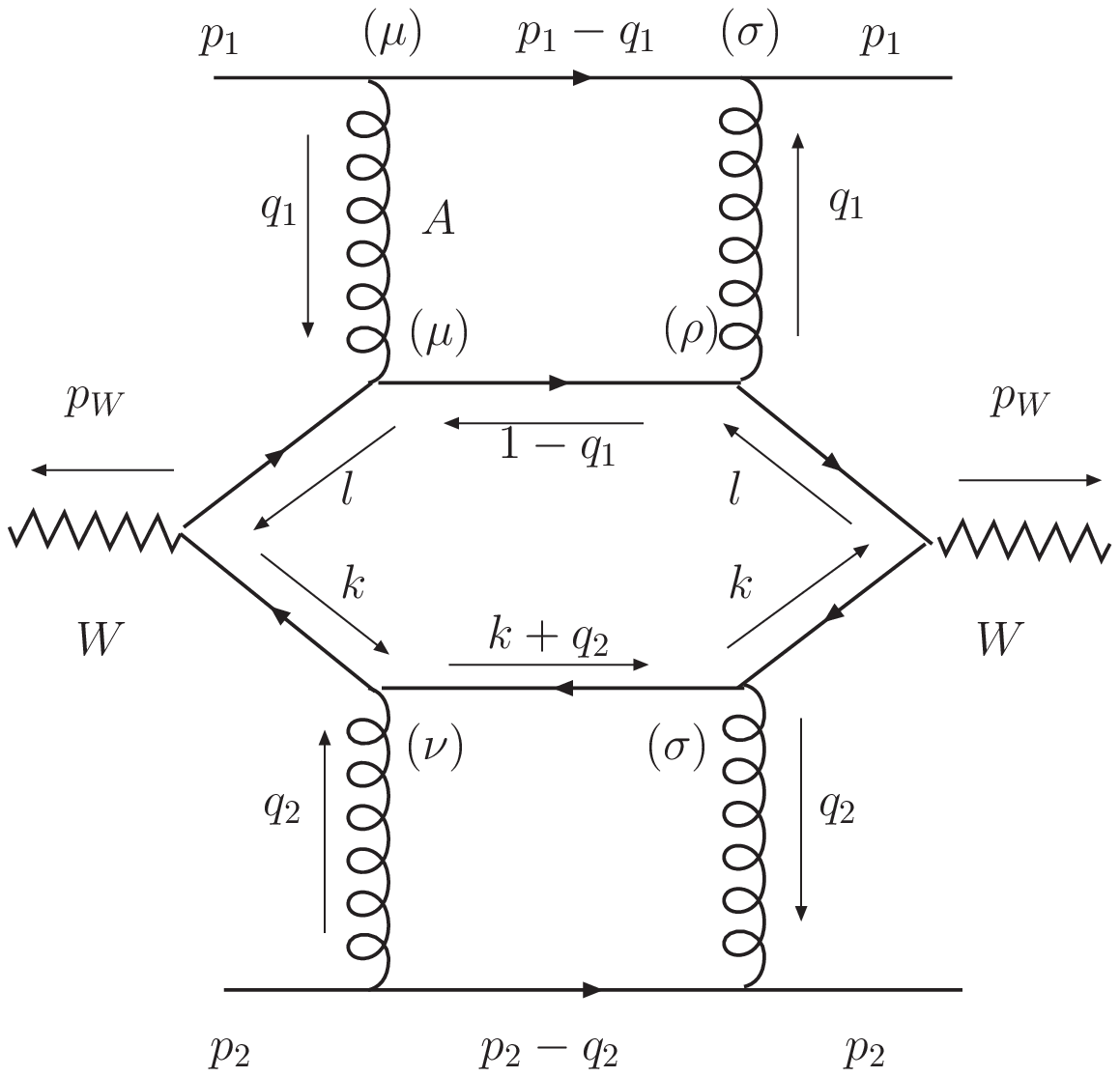,width=110mm,height=80mm}{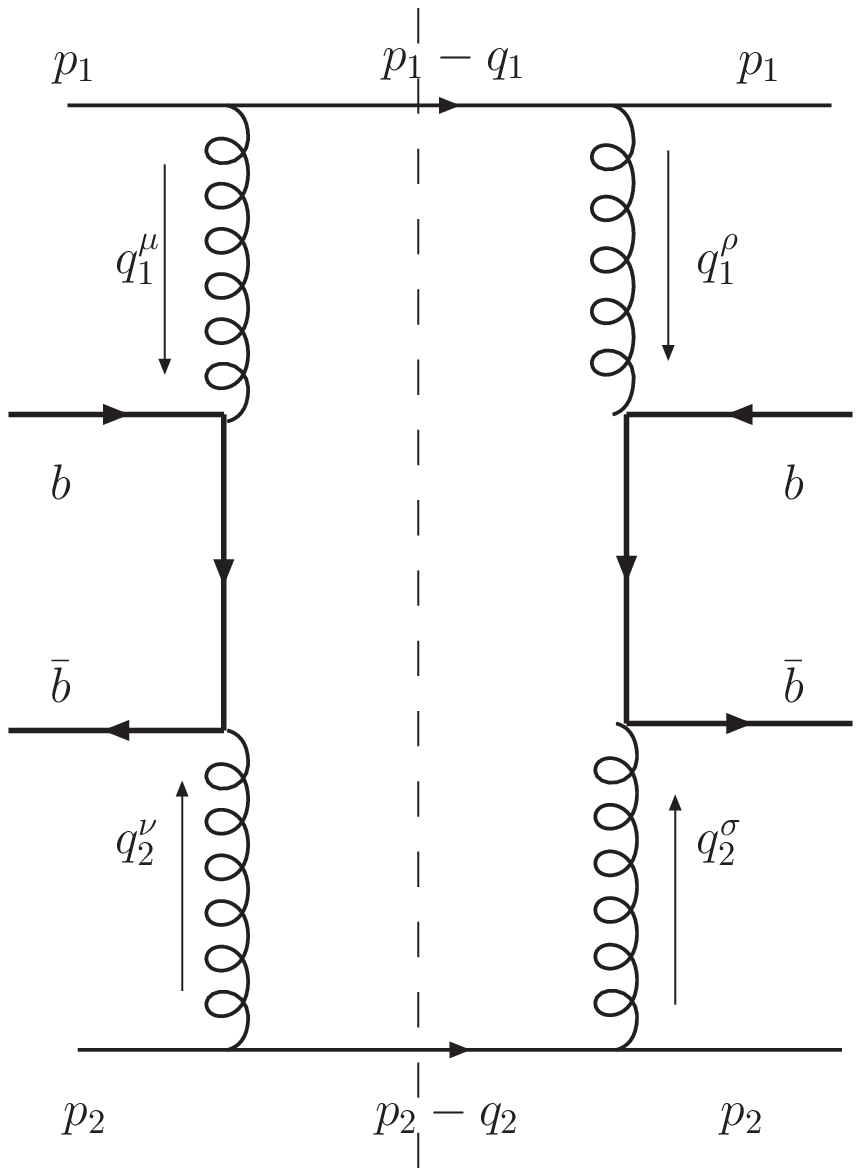,width=90mm,height=87mm}
{Born Approximation for W boson inclusive  production.
\label{hex}}{Born Approximation for quark anti-quark inclusive
production.\label{bbin} }

The Mueller diagram in the Born approximation of pQCD for the
inclusive W boson   production, is shown in \fig{incw},  and the
formula for this cross section is given by \eq{FT}.  We need only to
calculate $\sigma\Lb  M_{W}\Rb$ and $\sigma\Lb  M_{b \bar{b}}\Rb$.
The easiest way to derive these cross sections, is to calculate them
in the Born approximation of perturbative QCD ( see  \fig{hex} and
\fig{bbin}). The calculations of the expressions for the cross
sections  for the processes of \fig{hex} and \fig{bbin}, are
presented in appendices A-3 and A-4, and the final expressions for
the amplitudes  of these diagrams take the form

\begin{itemize}
\item \quad
  for $ \sigma\Lb M_W,q_1,q_2 \Rb$ (see \fig{hex} and \eq{Ad})
\bea \sigma(M_W,q_1,q_2)\,\,\,&=&\,\,
\,\,\frac{\as^2(q^2)}{q^2_{1,\perp}\,\,q^2_{2,\perp}}
\,\sigma\Lb M_W\Rb \,\,\,\,\,\, \notag\\
\mbox{where}\,\,\,\,\,\,\,\,\,\,\,\,\,\,\,\,\,\,\,\,\,\,\,\,\,\,\,\,\,\,\,
\sigma\Lb M_W\Rb &=& \Lb\frac{N^2_c - 1}{2 N_c}\Rb^2 \frac{1}{4\,
N_c}\frac{\as^2}{4
\,\pi^3}\,\sqrt{2}\,G_F\ln\Lb\frac{M_W^2}{4m_u^2}\Rb\ln\Lb\frac{M_W^2}{4m_d^2}\Rb
\label{WPR}\eea

 Recall that the Fermi coupling $G_F$ has the
dimensions of $\mbox{GeV}^{-2}$, so that inspection of \eq{WPR},
shows that $\sigma\Lb M_W\Rb$ has the dimensions of the cross
section. The factor in front of $\sigma\Lb M_W\Rb$, (in the first
line of \eq{WPR}), is taken into account in the BFKL Pomeron
exchange in \eq{FT}. Integrating over the transverse momenta
$\vec{q}_{1,\bot}$ and $\vec{q}_{2,\bot}$ on the RHS of \eq{WPR},
one obtains the following expression for the Born amplitude of
\fig{hex}.

\bea \sigma(M_W,q_1,q_2)\,\,\,=\,\,
\,\,\as^2(q^2)\pi^2\ln^2\Lb\frac{M_W^2}{q^2_{\bot,\mbox{\footnotesize{min}}}}\Rb
\,\sigma\Lb M_W\Rb \label{WPRI}\eea

\item \quad  for $ \sigma\Lb M_{b \bar{b}},q_1,q_2 \Rb$ (see \fig{bbin} and \eq{Ab})

\bea
 \sigma(M_{b
\bar{b}},q_1,q_2)\,\,\,
 &=&\,\,\,\,\frac{
\as^2(q^2)}{q^2_{1,\perp}\,\,q^2_{2,\perp}}
\,\sigma\Lb M_{b \bar{b}} \Rb \,\,\,\,\,\, \notag\\
\mbox{where}\,\,\,\,\,\,\,\,\,\,\,\,\,\,\,\,\,\,\,\,\,\,\,\,\,\,\,\,\,\,\,
\sigma\Lb M_{b \bar{b}}\Rb\, &=& \,\Lb\frac{N^2_c - 1}{2 N_c}\Rb^2
\frac{1}{4\, N_c}\frac{\as^2}{8 \,\pi}\,\,\frac{1}{M^2_{b
\bar{b}}}\ln\Lb\frac{M_W^2}{4m_b^2}\Rb \label{BBPR} \eea

  Integrating over the transverse momenta
$\vec{q}_{1,\bot}$ and $\vec{q}_{2,\bot}$ on the RHS of \eq{BBPR},
one obtains the following expression for the Born amplitude of
\fig{bbin}.

\bea \sigma(M_{b\bar{b}},q_1,q_2)\,\,\,=\,\,
\,\,\as^2(q^2)\pi^2\ln^2\Lb\frac{M_{b\bar{b}}^2}{q^2_{\bot,\mbox{\footnotesize{min}}}}\Rb
\,\sigma\Lb M_{b\bar{b}}\Rb \label{BBPRI}\eea

\end{itemize}

\numberwithin{equation}{section}
 \numberwithin{equation}{subsection}
\section{Two parton shower contribution to the background for W boson and Higgs production}
\subsection{The simplest diagram}

For the associate W boson and Higgs production, we expect a small
background, since the main process which creates such a background,
namely,

\beq \label{MBG} p + p \rightarrow  b \bar{b} + W + X \eeq

 leads to
a negligible contribution if both the $W$ and the $b \bar{b}$ pair
are produced from one parton shower, at $Y_{b \bar{b}} \approx Y_W$.
Here, $Y_{b \bar{b}}$ is the value for the rapidity of the quark
anti-quark pair, namely, $Y_{b \bar{b}} = \h( Y_b  + Y_{\bar{b}})$
and $Y_W$ is the Higgs boson rapidity.  Indeed, in one parton
shower, the typical difference ($\Delta y$)  in  the rapidity
between the two emitted partons, are larger (or equal to) $1/\as
\,\gg 1$. Therefore, keeping $ |Y_{b \bar{b}} - Y_W| \leq 1/\as$, we
have a suppression for the production of the W boson and the quark
anti-quark pair.

However, there exists a process for the production of the quark
anti-quark pair and W boson production, from two different parton
showers (see \fig{2ps}). In this process, we do not expect any
suppression of the quark anti-quark pair, and W boson production, at
the same values of rapidity. However, this has its own suppression,
which is related to the small probability associated with having two
parton shower processes. However, at high energies, this suppression
is not actually very strong, as we will demonstrate below.

\FIGURE[h]{ \centerline{\epsfig{file=
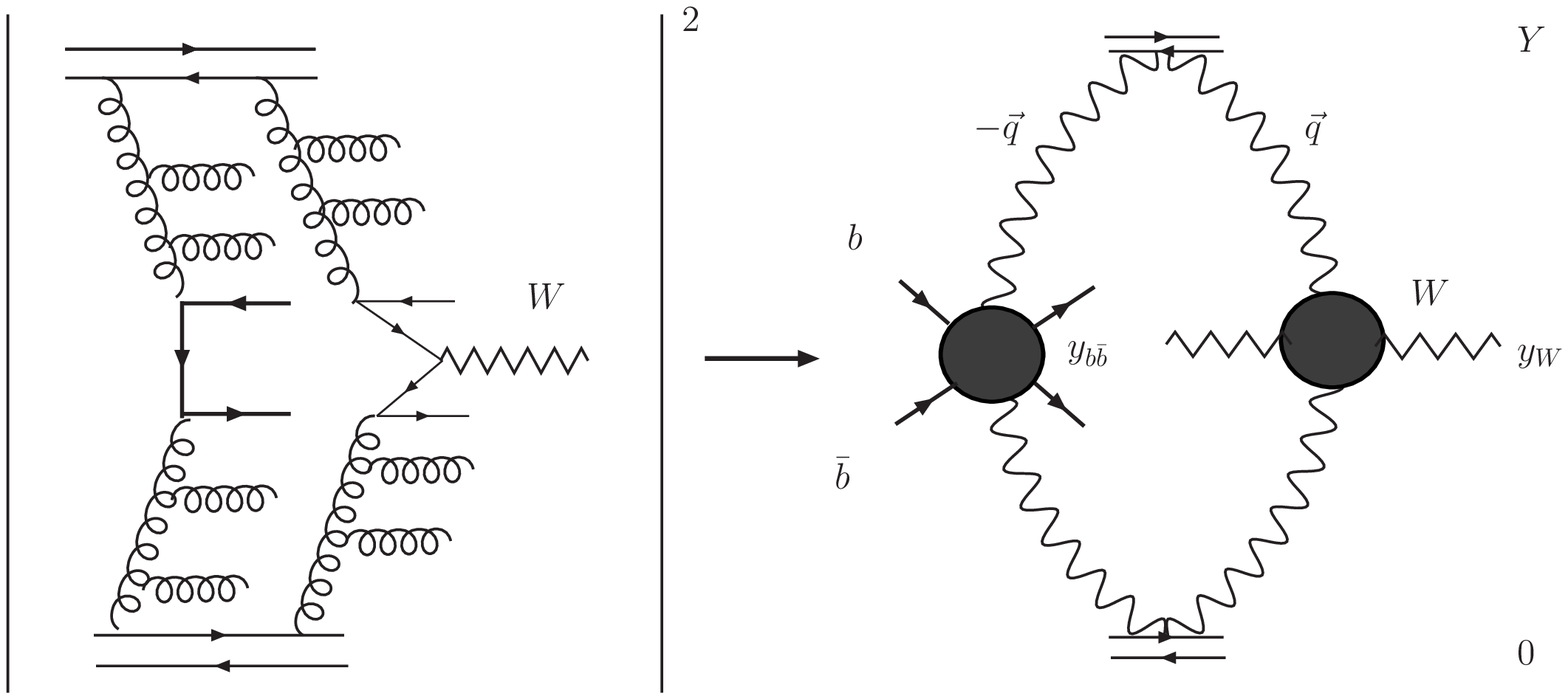,width=110mm,height=50mm}} \caption{The simplest Mueller
diagram for  the two  parton shower contribution to the background
for the W boson and Higgs production} \label{2ps}}

The general expression for the diagram of \fig{2ps} has the
following form

\beq \label{E1} A\Lb \fig{2ps} \Rb\,\,=\,\,\int\,\,\frac{d^2 q}{(2
\pi)^2}\,V^2\Lb pp \to \p\p; \vec{q} \Rb\,\, \frac{d
\sigma_{incl}\Lb\ln(s/M^2_{b , \bar{b}}); Y_{b \bar{b}}; - \vec{q}
\Rb}{d Y_{b \bar{b}}}\,\, \,\,\frac{d
\sigma_{incl}\Lb\ln(s/M^2_{W_{~}}); Y_{W_{~}};  -\vec{q} \Rb}{d
Y_{W}} \eeq

 $V\Lb pp \to \p\p\Rb$ in \eq{E1} and \eq{E2} below is
the vertex of the interaction of the two Pomerons with the proton
(see \fig{2ps}). In \eq{E1}, $d \sigma_{incl}/d y $ is the cross
section for the inclusive production of the W boson, and/or the
quark anti-quark pair, but not at $q=0$. However, at high energy and
for hard processes, we can assume that the value of the typical $q$
is small, and it is of the order of the typical momenta inside of
the proton. At high energy, the mechanism for inclusive production
is related to the emission of one parton shower, which can be
described by the DGLAP evolution \cite{DGLAP}, or by the BFKL
Pomeron exchange \cite{BFKL}. For both cases, it has been proved
\cite{GLR,BART,LMP} that we can safely put $q =0$ in our evaluation
of the integral of \eq{E1}. Therefore,

 \bea \label{E2}
 A\Lb \fig{2ps} \Rb\,\,&=&\,\,\frac{d
\sigma_{incl}\Lb \ln(s/M^2_{b , \bar{b}});Y_{b \bar{b}}; 0 \Rb}{d
Y_{ b\bar{b}}}
    \,\,\frac{d \sigma_{incl}\Lb \ln(s/M^2_W);Y_{W};0\Rb}{ Y_{W}}\,\times \,\,\int\,\,\frac{d^2 q}{(2 \pi)^2}\,V^2\Lb pp \to \p\p; \vec{q} \Rb \nonumber \\
 & = &\,\frac{\frac{d \sigma_{incl}\Lb\ln(s/M^2_{b , \bar{b}}); Y_{b \bar{b}}; 0 \Rb}{d Y_{b \bar{b}}}
\,\,\frac{d \sigma_{incl}\Lb\ln(s/M^2_W); Y_W; 0 \Rb}{d Y_W}}{2
\sigma_{eff}} \eea

 Fortunately for us, the value of $\sigma_{eff}$ has
been measured by the CDF collaboration at the Tevatron \cite{DPXS},
with the result $\sigma_{eff}\,\,=\,\,14.5 \pm 1.7 \pm
2.3\,\mbox{mb}$. Using the value of the predicted cross section for
$b \bar{b}$ production, taking $M_{b\bar{b} }= M_{H} = 100\,\, \GVc$
\cite{BBXS} for $Y_{b \bar{b}}$ = 0, and at the LHC energy, namely,
$\frac{d \sigma_{incl}\Lb\ln(s/M^2_{b , \bar{b}}); Y_{b \bar{b}}; 0
\Rb}{d Y_{b \bar{b}}}\,\,=\,\,2\,\times\,10^{-6}\,
\mbox{mb}$\footnote{This value, together with a more detailed
consideration of the value of the cross section for $b \bar{b}$
inclusive production, will be discussed in section 4}, one can
obtain from \eq{E2} (comparing it with \eq{WH2W}), that

 \beq \label{E3} \frac{d
\sigma_{background}\Lb \eq{E2} \Rb / d y_{b \bar{b}}|_{Y_{b \bar{b}}
=0}}{ d \sigma_{H\,\,\,production}/d
Y_H|_{Y_H=0}}\,\,\approx\,\,2\,\,\,\,\,\mbox{in the LHC energy
range} \eeq

 This ratio can be easily reduced using the
characteristic property of the process of \fig{2ps}, namely that
there are no correlations in the rapidity values between the $W$
boson, and the quark anti-quark pair. In other words,  the cross
section for this process does not depend on the difference $ Y_W -
Y_{b  \bar{b}}$, while the cross section for the associate W Higgs
production of \fig{hwprod},  has a maximum at $ Y_W - Y_{b  \bar{b}}
= 0$.

This result is encouraging, but we need to estimate the more
complicated diagrams to derive the final conclusions (see
\fig{2psen}).

\subsection{The two parton shower production: enhanced diagram}
In the diagram of \fig{2psen}, the two parton showers are produced
at the rapidity $y_1$,  and they merge back to one parton shower at
the rapidity $y_2$. Such a process can be reduced to the enhanced
BFKL Pomeron diagram at high energy, as it is shown in \fig{2psen}.
We would like to recall, that in this simplification, we use the
unitarity constraint

\beq \label{UNCON} 2 \mbox{Im}\,A_{el}\,\,\,=\,\,| A_{el}|^2
\,\,+\,\,G_{in} \eeq

 where $A_{el}$ is the elastic scattering
amplitude at energy $s$ and impact parameter $b$, while $G_{in}$ is
the contribution of all inelastic processes.  The scattering
amplitude is purely imaginary at high energy, and it can be
described by the exchange of one BFKL Pomeron. Neglecting $|A_{el}|$
in \eq{UNCON}, which is possible  for the range of energies  $\as
\ln (s/s_0)\,\ll\,\ln(1/\as^2)$ , we obtain from \eq{UNCON} the
relation

 \beq \label{UNCON1} 2\, A\Lb 1\p\Rb\,\,=\,\,G_{in}(s,b)
\eeq

 where $A\Lb 1\p\Rb$ is the amplitude for one BFKL Pomeron
exchange.  Using this relation, we are able to replace the
production of the parton showers with the exchange of the BFKL
Pomerons. In this subsection, we will calculate the simplest
enhanced diagram shown in \fig{2psen}, whose contribution is given
by the expression

 \bea
&&A\Lb \p \to 2\p \to \p;\fig{2psen} \Rb =\,\oint_C \!
\,h\Lb\,\gamma\Rb\,\,\, d \gamma
\,\prod^2_{i=1}\oint_{C_i}\,h\Lb\,\gamma_i\Rb\,\,\,d
\gamma_i\,\oint_{C_i\,'}\,h\Lb\,\gamma'_i\Rb\,\,
d \gamma'_i\,\oint_{C\,'}\,h\Lb\,\gamma'\Rb\,d \gamma'\,\nonumber\\
&&\times\,\, \,\,\,\,\, \,\int^{Y}_{Y_W +\h \delta
Y_W}\!\!\!\!\!\!\!dY_1\int^{Y_{b \bar{b}} - \h \delta Y_{b
\bar{b}}}_{0}\!\!\!\!\!\!\!dY_2\,\, g_P\Lb \vec{k}, \vec{q}=0;
-\gamma \Rb\,\lambda^{-1}\Lb\,\gamma\Rb \,e^{\omega(\gamma)\,(Y -
Y_1)}\,\,\,
  \, e^{(\omega(\gamma_1)\,( Y_1 - Y_W - \h \delta Y_W)} \nonumber\\
&&\times\,\,\, e^{\omega(\gamma_2)\,( Y_1 - Y_{b \bar{b}} - \h
\delta Y_{b \bar{b}})}
  \,\,\, e^{\omega(\gamma'_1)\,(Y_W - \h \delta Y_W - Y_2)}\, e^{\omega(\gamma'_2)\,(Y_{b \bar{b}} - \h \delta Y_{b \bar{b}} - Y_2)}\,e^{\omega(\gamma')\,(Y_2 - 0)}\,\nonumber\\
  &&\times\,g_P\Lb \vec{k}_0, \vec{q}=0 ;\gamma'  \Rb\,\lambda^{-1}\Lb\,\gamma'\Rb\,\,\sigma(M_{W},-\gamma_1,\gamma'_1)\, \sigma(M_{b \bar{b}},-\gamma_2,\gamma'_2)\notag\\
  &&\times\, \int\!d^2 q\,'\Gamma_{3\p} \Lb \vec{q}=0, - \vec{q}\,'| \gamma,\gamma_1,\gamma_2 \Rb \,\,
\Gamma_{3\p} \Lb \vec{q}=0, - \vec{q}\,'
|-\gamma',-\gamma'_1,-\gamma'_2 \Rb
 \label{FED1g}
\eea

\FIGURE[h]{
\centerline{\epsfig{file=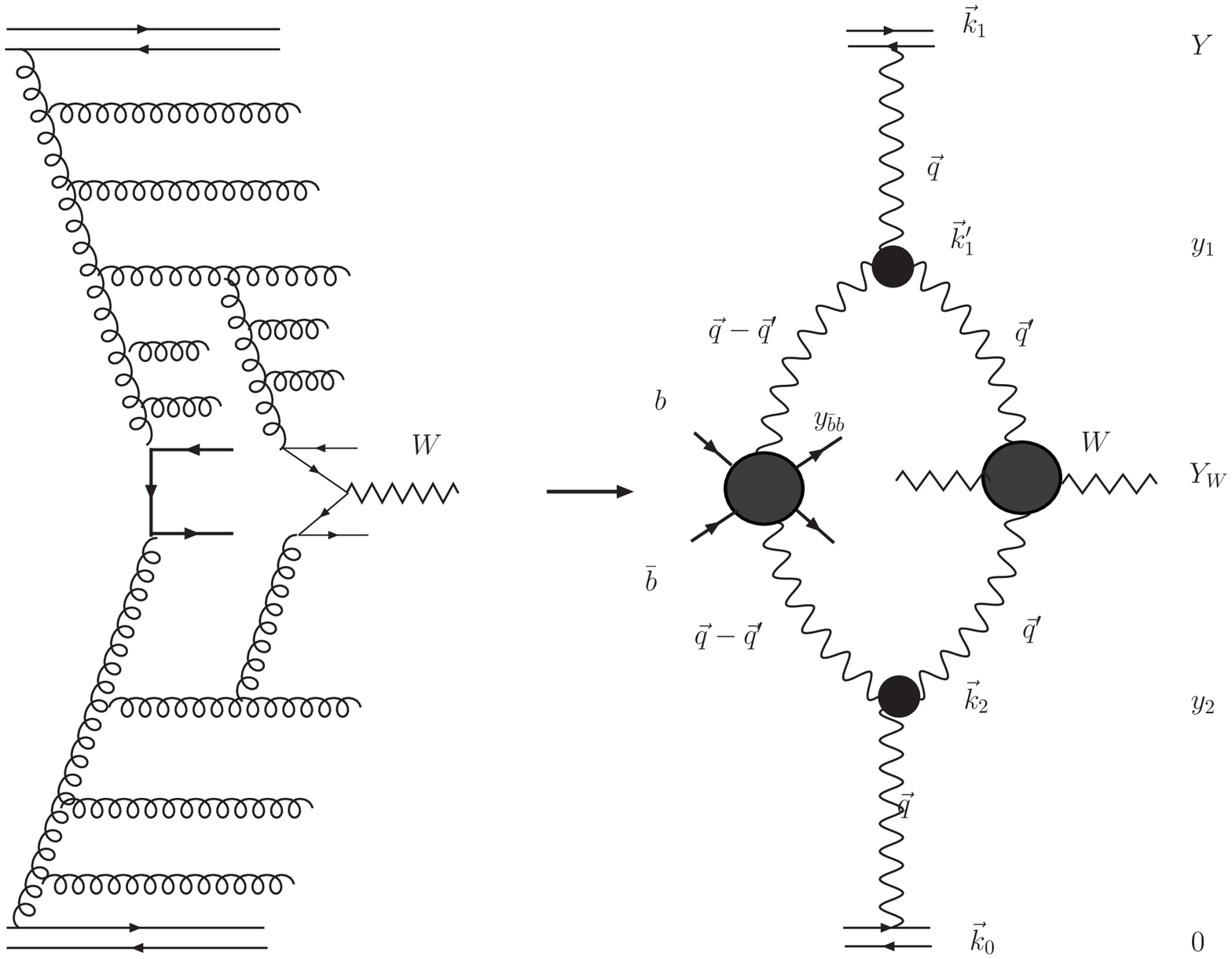,width=100mm,height=65mm}}
\caption{The first enhanced  Mueller diagram for  two  parton shower
contribution to the background for the W boson and Higgs production}
\label{2psen}}

where $h\Lb\,\gamma\,\Rb\,$ and
$\lambda\Lb\,\gamma\Rb\,=\,\Lb\,\gamma\Lb\,\gamma-1\Rb\,\Rb^{-2}$
have been defined in \eq{1P1g}. The contours $C$, $C'$ , $C_1$ ,
$C_2$ , $C_1'$ and $C_2'$ are similar to the contour shown in
\fig{graph}, which consists of the imaginary $\gamma$ axis from
$\pm\,i\infty$, and the semi circle at $\infty$ to the left of the
imaginary $\gamma$ axis, which encloses all singularities in the
integrand of \eq{FED1g}. In this case also, it is assumed that the
integrand as a function of $\gamma$, $\gamma'$, $\gamma_1$,
$\gamma_2$, $\gamma_1'$ and $\gamma_2'$, vanishes on the semi circle
at $\infty$, such that it is sufficient just to replace for each
integration over the conformal variables $\gamma_l$, in \eq{FED1g},

\beq
\oint_C\,d\gamma_l\,\rightarrow\,\int^{\epsilon\,+i\infty}_{\epsilon-\,i\,\infty}d\gamma_l\,\,\,\,\,\,\,\,\,\,\,\,\,\,\,\,\,\,\,\,\{\gamma_l\}
= \gamma, \gamma',\gamma_1, \gamma_2, \gamma'_1,
\gamma'_2\label{rep12} \eeq

Following the definition of \eq{gamdef}, one can replace the
integration limits for the $\gamma_l$ variables
$\epsilon-i\infty\leq\gamma_l\leq\epsilon+i\,\infty$, with the
corresponding limits of integration for the variables $\nu_l$
($\{\nu_l\} = \nu, \nu',\nu_1, \nu_2, \nu'_1, \nu'_2$), namely
$-\infty\leq\nu_l\leq\infty$. So in \eq{FED1g}, one can change from
the variable $\gamma_l$ to the variable $\nu_l$, and re-express the
integration over $\gamma_l$ as

\beq \label{notations}
\oint_{C}\,d\gamma_l\,\rightarrow\,\int^{\epsilon\,+i\infty}_{\epsilon-\,i\,\infty}d\gamma_l\,\,=\,\,\int^{
i a + \infty}_{i a - \infty}\,d\nu_l,\,\,\,\,\,\,\,\,\{\nu_l\} =
\nu, \nu',\nu_1, \nu_2, \nu'_1, \nu'_2 \eeq

 where $a\,=\,1/2-\epsilon\,$ is such that
the contour of integration is below all singularities in $\nu_l$.
Hence, in $\nu$ notation, \eq{FED1g} will be rewritten as follows,
where the notation $\intf\,d\nu$ is understood to be a shorthand for
the contour integrals written in \eq{notations}.

 \bea
&&A\Lb \p \to 2\p \to \p;\fig{2psen} \Rb =\,\pItnprr\,\int\,d^2 q\,'\,\, \nonumber \\
&&\times  \,\int^{Y}_{Y_W + \h \delta
Y_W}\!\!\!\!\!\!\!dY_1\int^{Y_{b \bar{b}} - \h \delta Y_{b
\bar{b}}}_{0}\!\!\!\!\!\!\!dY_2\,\, g_P\Lb \vec{k}, \vec{q}=0; -\nu
\Rb \,e^{\omega(\nu)\,(Y - Y_1)}\,\,
  \, e^{\omega(\nu_1)\,( Y_1 - Y_W - \h \delta Y_W)} \,\, e^{\omega(\nu_2)\,( Y_1 - Y_{b \bar{b}} - \h \delta Y_{b \bar{b}})} \nonumber \\
&& \times \,\,\, e^{\omega(\nu'_1)\,(Y_W - \h \delta Y_W - Y_2)}\, e^{\omega(\nu'_2)\,(Y_{b \bar{b}} - \h \delta Y_{b \bar{b}} - Y_2)}\,e^{\omega(\nu')\,(Y_2 - 0)}\,g_P\Lb \vec{k}_0, \vec{q}=0 ;\nu'  \Rb\,\,\,\sigma(M_{W},-\nu_1,\nu'_1)\, \sigma(M_{b \bar{b}},-\nu_2,\nu'_2) \nonumber \\
&&\times \,\,\,\Gamma_{3\p} \Lb \vec{q}=0, - \vec{q}\,'|
\nu,\nu_1,\nu_2 \Rb \,\, \Gamma_{3\p} \Lb \vec{q}=0, - \vec{q}\,'\,
|-\nu',-\nu'_1,-\nu'_2 \Rb
 \label{FED1}
\eea

\bea \mbox{where}\,\,\,\,\,\,\,\,\,\,\,\,\,\,\,\,\,\,
\sigma(M_W,-\nu_1,\nu'_1)\,\,\,&=&\,\,\,g(M_W,\vec{q}=0;-\nu_1)\,\,g(M_W,\vec{q}=0;\nu'_1)\,\sigma(M_W)\notag\\
\mbox{and}\,\,\,\,\,\,\,\,\,\,\,\,\,\,\,\,\,\,\sigma(M_{b\bar{b}},-\nu_2,\nu'_2)\,\,\,&=&\,\,\,g(M_{b\bar{b}},\vec{q}=0;-\nu_2)\,\,g(M_{b\bar{b}},\vec{q}=0;\nu'_2)\,\sigma(M_{b\bar{b}})
\label{sigma} \eea

 where the expressions for the $\sigma\Lb M_W\Rb$ and $\sigma\Lb M_{b\bar{b}}\Rb$ are
given in \eq{WPR} and \eq{BBPR}, respectively. In \eq{FED1}, we
assume that $Y_W
> Y_{b \bar{b}}$, and that $q=0$, since we are considering the inclusive process
given by \fig{2psen}, (see the left part of this figure). From
\eq{FED1}, we see that it is possible to integrate over $q\,'$. This
integration has been performed in Ref. \cite{LMP}, with the result

\bea && \Gamma_{3\p} \Lb q=0,q'| \nu,\nu_1,\nu_2 \Rb
 \equiv \NA\int\,d^2 k\,\,\,g\Lb k,q=0,\nu \Rb \,\,g\Lb  \kv + \frac{1}{2} \qv\,', \qv\,' , \nu_1\Rb \,\,
g\Lb  \kv + \frac{1}{2}  \qv\,',- \qv\,' ,\nu_2\Rb  \notag\\
&&\xrightarrow{\nu_1 \to 0,\, \nu_2 \to 0}
\,\,\NA\frac{2^{1+2i\nu}C_1(\nu)\, \pi^4\, \Lb
\frac{q'^2}{4}\Rb^{-1/2 + i \nu - i \nu_1 - i \nu_2}}{(1/2 + i \nu -
i \nu_1 - i \nu_2)^3}\,\,+\,\,\mbox{function
without singularities in $\nu_1$ and $\nu_2$}  \nonumber \\
\label{G3P}\eea

In \eq{G3P},  we restrict ourselves to the case of small $\nu_1$ and
$\nu_2$ ( $\nu'_1$ and $\nu'_2$),  since as we will show below, the
main contribution to the integral over rapidity values, stems from
$Y_1 - Y_2 \to Y$, and small values of $\nu_1$ and $\nu_2$  (
$\nu'_1$ and $\nu'_2$),   contribute to the BFKL Pomeron exchange at
high energy (see \eq{OMe}, and the discussion around this equation).
 From \eq{G3P}, the result of integrating the product of the two
triple pomeron vertices, which appear in the integrand on the RHS of
\eq{FED1}, is

\bea \label{GG}
&&\int\!d^2q\,'\Gamma_{3\p}\Lb\,\vec{q}=0,\vec{q}\,'|\nu\,,\nu_1\,,\nu_2\Rb\,\Gamma_{3\p}\Lb\,\vec{q}=0,\vec{q}\,'|-\nu'\,,-\nu'_1\,,-\nu'_2\Rb\,\,=\\
&&\,\NA^2\frac{2^{2i\nu-2i\nu'}C_1\Lb\,\nu\Rb\,C_1\Lb\,-\nu'\Rb\,2\pi^92^{\Lb\,2i\nu-2i\nu_1-2i\nu_2-2i\nu'+2i\nu_1'+2i\nu_2'\Rb\,}}{\Lb\,1/2+i\nu-i\nu_1-i\nu_2\Rb^3\,\Lb\,1/2-i\nu'+i\nu_1'+i\nu_2'\Rb^3\,}\,\,
\times\,\,\delta\Lb\,\nu-\nu_1-\nu_2-\nu'+\nu_1'+\nu_2'\Rb\,\notag
\eea

The Dirac delta function appearing on the RHS of \eq{GG}, is
absorbed in \eq{FED1} by integrating over $\nu'$. In the course of
this, the
expression on the RHS of \eq{GG}  reduces to\\
$\NA^2\,C_1\Lb\,\nu\Rb\,C_1\Lb\,\nu-\nu_1-\nu_2+\nu_1'+\nu_2'\Rb\,2\pi^9\Lb\,1/2+i\nu-i\nu_1-i\nu_2\Rb^{-3}\,\Lb\,1/2+i\nu-i\nu_1-i\nu_2\Rb^{-3}\,$.\\
Hence, after evaluating the integration over $\nu'$, and making use
of the expression in \eq{Ge0}, in the case where $\vec{q}=0$ for the
pomeron couplings, gives for the RHS of \eq{FED1}

\bea &&A\Lb \p \to 2\p \to \p;\fig{2psen}
\Rb\,=\,2\pi^9\NA^2\Gamma_P\Lb k,q-0\Rb\Gamma_P\Lb
k_0,q=0\Rb\sigma\Lb
M_W\Rb\sigma\Lb M_{b\bar{b}}\Rb\notag\\
\,\,\, &&\times\,\pItn\, \Lb\,\nu_1'+\nu_2'-\nu_1-\nu_2+\nu\Rb^2
\, \, \,\,\,\nonumber \\
&&\times  \,\int^{Y}_{Y_W + \h \delta
Y_W}\!\!\!\!\!\!\!dY_1\int^{Y_{b \bar{b}} - \h \delta Y_{b
\bar{b}}}_{0}\!\!\!\!\!\!\!dY_2\,\, g\Lb \vec{k}, \vec{q}=0; -\,\nu
\Rb \,e^{\omega(\nu)\,(Y - Y_1)}\,\,
  \, e^{(\omega(\nu_1)\,( Y_1 - Y_W - \h \delta Y_W)} \,\, e^{\omega(\nu_2)\,( Y_1 - Y_{b \bar{b}} - \h \delta Y_{b \bar{b}})} \nonumber \\
&& \times \,\,\, e^{\omega(\nu'_1)\,(Y_W - \h \delta Y_W - Y_2)}\,
e^{\omega(\nu'_2)\,(Y_{b \bar{b}} - \h \delta Y_{b \bar{b}} -
Y_2)}\,e^{\omega(\nu-\nu_1-\nu_2+\nu_1'+\nu_2')\,(Y_2 - 0)}\,\,\,
\notag\\
&&\times\,g(M_{W},-\nu_1)\,g(M_{W},\nu'_1) \, g(M_{b\bar{b}},-\nu_2)\,g(M_{b\bar{b}},\nu'_2) \nonumber \\
&&\times
\,\,\,C_1\Lb\,\nu\Rb\,C_1\Lb\,\nu_1+\nu_2-\nu_1'-\nu_2'-\nu\Rb\,\Lb\,1/2-i\nu+i\nu_1+i\nu_2\Rb^{-3}\,\Lb\,1/2+i\nu-i\nu_1-i\nu_2\Rb^{-3}\,\notag\\
&\times\,&g\Lb \vec{k}_0, \vec{q}=0 ;\nu_1'+\nu_2'-\nu_1-\nu_2+\nu
\Rb\,
 \label{FED2}
 \eea

where from the expression of \eq{Ge0}, the product $C_1\,C_1 g \,g$
can be recast in the form

\bea \label{CCGG} &&
C_1\Lb\,\nu\Rb\,C_1\Lb\,-\nu_1'-\nu_2'+\nu_1+\nu_2-\nu\Rb\,g\Lb
\vec{k}, \vec{q}=0; -\nu \Rb\,g\Lb \vec{k}_0, \vec{q}=0
;\nu_1'+\nu_2'-\nu_1-\nu_2+\nu \Rb\,  \\
&&\longrightarrow \,\,
\,\,C_1\Lb\,\nu\Rb\,C_1\Lb\,-\nu\Rb\,C_1\Lb\,\nu_1'+\nu_2'-\nu_1-\nu_2+\nu\Rb\,C_1\Lb\,-\nu_1'-\nu_2'+\nu_1+\nu_2-\nu\Rb\ \nonumber   \\
&&\Lb\,k^2\,k_0^2\Rb^{-1/2}\Lb\,16\,k^2\Rb^{-i\nu}
\,\,\Lb\,16\,k_0^2\Rb^{i\nu_1+i\nu_2-i\nu_1'-i\nu_2'+i\nu} \nonumber
\eea

Hence, it is obvious that using the relation of \eq{CC}, the
expression in \eq{CCGG} cancels the term
$\nu^2\,\Lb\,\nu_1'+\nu_2'-\nu_1-\nu_2+\nu\Rb^2$ in the numerator of
the integrand of \eq{FED2}, to give

 \bea
&&A\Lb \p \to 2\p \to \p;\fig{2psen} \Rb
=\,\pi^8\,\Sigma\Lb k,M_W,M_{b \bar{b}},k_0 \Rb \notag\\
&&\times\,\Itn\,\prod^2_{i=1}\Itn_i\nu^2_i\,\lambda\Lb\,\nu_i\Rb\,\,
\,\Itnpr_i\nu'^2_i\,\lambda(\nu'_i)\, \, \, \Lb\,16\,k^2\Rb^{-i\nu}
\,\,\Lb\,16\,k_0^2\Rb^{i\nu_1+i\nu_2-i\nu_1'-i\nu_2'+i\nu}\,\,\,\nonumber \\
&&\times  \,\int^{Y}_{Y_W + \h \delta
Y_W}\!\!\!\!\!\!\!dY_1\int^{Y_{b \bar{b}} - \h \delta Y_{b
\bar{b}}}_{0}\!\!\!\!\!\!\!dY_2\,\,  \,e^{\omega(\nu)\,(Y -
Y_1)}\,\,
  \, e^{\omega(\nu_1)\,( Y_1 - Y_W - \h \delta Y_W)} \,\, e^{\omega(\nu_2)\,( Y_1 - Y_{b \bar{b}} - \h \delta Y_{b \bar{b}})}  \label{FED3'} \\
&& \times \,\,\, e^{\omega(\nu'_1)\,(Y_W - \h \delta Y_W - Y_2)}\, e^{\omega(\nu'_2)\,(Y_{b \bar{b}} - \h \delta Y_{b \bar{b}} - Y_2)}\,e^{\omega(\nu_1+\nu_2-\nu_1'-\nu_2'+\nu)\,(Y_2 - 0)}\,\,\, \nonumber \\
&&\times\,g(M_{W},-\nu_1)\,g(M_{W},\nu'_1) \,
g(M_{b\bar{b}},-\nu_2)\,g(M_{b\bar{b}},\nu'_2)
\,\,\,\Lb\,1/2-i\nu+i\nu_1+i\nu_2\Rb^{-3}\,\Lb\,1/2+i\nu-i\nu_1-i\nu_2\Rb^{-3}\,
\notag
 \eea

 where the  factor

\beq \label{SIDE} \Sigma \Lb k,M_W,M_{b \bar{b}},k_0
\Rb\,\,=\,\,2\,\pi^9\,\,\Lb \frac{2 \pi \bas}{N_c}\Rb^2\,\Gamma_P\Lb
k,q=0\Rb\,\Gamma_P\Lb k_0,q=0\Rb\,\, \frac{1}{\sqrt{k^2\,k^2_0}}\,\,
\sigma\Lb M_{b \bar{b}}\Rb\,M^2_{b\bar{b}} \sigma\Lb
M_W\Rb\,M_W^2\eeq

Substituting for $g$ appearing in \eq{FED3'} for the form given in
\eq{G},  the expression on the RHS of \eq{FED3'} becomes

  \bea
&&A\Lb \p \to 2\p \to \p;\fig{2psen} \Rb\,=\,\pi^{16}\,\Sigma\Lb k,M_W,M_{b \bar{b}},k_0 \Rb\notag\\
&&\times\,\Itn\,\prod^2_{i=1}\Itn_i\nu_i\,\lambda\Lb\,\nu_i\Rb\,\,
\,\Itnpr_i\nu'_i\,\lambda(\nu'_i)\, \, \,\Lb\,16\,k^2\Rb^{-i\nu}
\,\,\Lb\,16\,k_0^2\Rb^{i\nu_1+i\nu_2-i\nu_1'-i\nu_2'+i\nu} \,\,\,\nonumber \\
&&\times  \,\int^{Y}_{Y_W + \h \delta
Y_W}\!\!\!\!\!\!\!dY_1\int^{Y_{b \bar{b}} - \h \delta Y_{b
\bar{b}}}_{0}\!\!\!\!\!\!\!dY_2\,\,  \,e^{\omega(\nu)\,(Y -
Y_1)}\,\,
  \, e^{\omega(\nu_1)\,( Y_1 - Y_W - \h \delta Y_W)} \,\, e^{\omega(\nu_2)\,( Y_1 - Y_{b \bar{b}} - \h \delta Y_{b \bar{b}})} \nonumber \\
&& \times \,\,\, e^{\omega(\nu'_1)\,(Y_W - \h \delta Y_W - Y_2)}\, e^{\omega(\nu'_2)\,(Y_{b \bar{b}} - \h \delta Y_{b \bar{b}} - Y_2)}\,e^{\omega(\nu_1'+\nu_2'-\nu_1-\nu_2+\nu)\,(Y_2 - 0)}\,\Lb\,16M^2_W\Rb^{i\nu_1'-i\nu_1}\Lb\,16M^2_{b\bar{b}}\Rb^{i\nu_2'-i\nu_2}\,\, \nonumber \\
&&\times
\,\,\,\Lb\,1/2-i\nu+i\nu_1+i\nu_2\Rb^{-3}\,\Lb\,1/2+i\nu-i\nu_1-i\nu_2\Rb^{-3}\,
 \label{FED311}
 \eea

Note that due to the $\nu_i$ and $\nu_i'$ ($i=\,1\,,\,2$) in the
denominator of \eq{G}, inserting this expression for the g's into
\eq{FED3'}, the $\nu_i^2$ and $\nu_i'^2$ in the numerator of
\eq{FED3'} reduce to just factors of $\nu_i$ and $\nu_i'$ in the
numerator of \eq{FED3'}. Since $\omega(\nu_l)\,>\,0$ contains a
minimum at $\nu_l = 0$,
 the integration over $Y_1$ and $Y_2$ is actually  convergent,
   either at $ Y - Y_1 \leq 1/\omega(0)$ ( $Y_2 \leq /\omega(0)$) or
   at $ Y_1 - Y_W  - \h \delta Y_W \leq 1/\omega(0)$
( $ Y_{b \bar{b}}  - \h \delta Y_{b \bar{b}}   - Y_2 \leq
1/\omega(0)$).   The maximal increase with energy stems from the
region where $Y_1 \to Y$ and $Y_2 \to 0$. Indeed, in this region the
 integrand in \eq{FED3'} is proportional to

 \beq \label{FED31a}
   e^{\omega(\nu_1)\,( Y - Y_W - \h \delta Y_W)} \,\, e^{\omega(\nu_2)\,( Y - Y_{b \bar{b}} - \h \delta Y_{b \bar{b}})}\,
 \times \, e^{\omega(\nu'_1)\,(Y_W - \h \delta Y_W )}\, e^{\omega(\nu'_2)\,(Y_{b \bar{b}} - \h \delta Y_{b \bar{b}} )}\,
\eeq

Using  \eq{FED31a}, and evaluating the integrations over all the
$\nu$'s in the saddle point approximation, we see that

 \beq \label{FED32} A\Lb \p
\to 2\p \to \p;\fig{2psen} \Rb\,\,\propto\,\exp \Lb \omega(0)\, \Lb
2 Y - \h \delta Y_W - \h \delta Y_{b \bar{b}} \Rb\,\,\Rb \eeq

 where the
typical values of $\nu_l$ are

 \bea \label{FED33}
|\nu_1|\,\approx\,\frac{\ln(M^2_W/k_0^2)}{2 ( Y - Y_W - \h \delta
Y_W)}\,\ll\,1;&\,\,\,\,\,\,&
|\nu_2|\,\approx\,\frac{\ln(M^2_{b \bar{b}}/k_0^2)}{2 ( Y - Y_{b \bar{b}} - \h \delta Y_{b \bar{b}})}\,\ll\,1;\,\,\, \nonumber  \\
|\nu'_1|\,\approx\,\frac{\ln(M^2_W/k^2_0)}{2 (  Y_W - \h \delta
Y_W)}\,\ll\,1;&\,\,\,& |\nu'_2|\,\approx\,\frac{\ln(M^2_{b
\bar{b}}/k^2_0)}{2 (  Y_{b \bar{b}} - \h \delta Y_{b
\bar{b}})}\,\ll\,1; \eea

 This can be seen more clearly in the argument that follows,
where considering the region of integration close to the point
$\nu\to\,0$, one obtains the delta function
$\delta\,\Lb\,Y-Y_1+Y_2\Rb\,$ in the rapidity variables.

For small $\nu_1$ and $\nu_2$  (  $\nu'_1$ and $\nu'_2$),  we have
three different kinematic regions of integration: (i) $|i \nu + 1/2|
\ll \nu_1 + \nu_2$ ($ |i \nu' + 1/2| \ll \nu'_1 + \nu'_2$);
 (ii) $|i \nu + 1/2| \gg \nu_1 + \nu_2$ ($ |i \nu' + 1/2| \gg \nu'_1 + \nu'_2$); and (iii) $ |i \nu + 1/2| \approx \nu_1 + \nu_2$ ($ |i \nu' + 1/2| \approx\nu'_1 + \nu'_2$).

\numberwithin{equation}{subsubsection} \setcounter{equation}{0}
\begin{boldmath}
\subsubsection{Region (I):
 $|i \nu + 1/2| \ll \nu_1 + \nu_2$ ($ |i \nu' + 1/2| \ll \nu'_1 +
\nu'_2$)}
\end{boldmath}



In the first kinematic region, we have
  $\Gamma_{3\p} \to 1/(  i \nu_1+i\nu_2 )^3$. Therefore, the integral is large for  values of $|\nu| \to 1/2$. Using \eq{GG},
 the  integrand in \eq{FED1} could be simplified in this kinematic
 region in the following way.

 \bea
&&A_{\mbox{reg}. (i)}\Lb \p
\to 2\p  \to \p;\fig{2psen} \Rb\,\, \,\,=\,\sigmaf\,\label{FED21}\\
&&\,\times\, \int^{\infty}_{-\infty} \! \,\nu^2\,\, d \nu\,
\,C_1(\nu) \,\prod^2_{i=1}\int^{\infty}_{-\infty}\!\nu^2_i\,\, d
\nu_i\,\int^{\infty}_{-\infty}\!\nu'^2_i \, d \nu'_i
\,\,\int^{\infty}_{-\infty}\!\nu'^2d\nu'C_1(-\nu')
\,\frac{\delta(\nu_1 + \nu_2 - \nu'_1 -
\nu'_2)2^{2i\nu_1+2i\nu_2-2i\nu_1'-2i\nu_2'}}{(\nu_1 +
\nu_2)^3\,(\nu'_1 + \nu'_2)^3}\,\,
 \nonumber  \\
&&\times \,\, g\Lb \vec{k}, \vec{q}=0; -\nu  \Rb \,\int^{Y}_{Y_W +
\h \delta Y_W}\!\!\!\!\!\!\!dY_1\int^{Y_{b \bar{b}} - \h \delta Y_{b
\bar{b}}}_{0}\!\!\!\!\!\!\!dY_2\,\,\,e^{\omega(\nu)\,(Y - Y_1)}\,\,
  \, e^{(\omega(\nu_1)\,( Y_1 - Y_W - \h \delta Y_W)} \,\, e^{\omega(\nu_2)\,( Y_1 - Y_{b \bar{b}} - \h \delta Y_{b \bar{b}})} \nonumber \\
&& \times \,\,\, e^{\omega(\nu'_1)\,(Y_W - \h \delta Y_W - Y_2)}\,
e^{\omega(\nu'_2)\,(Y_{b \bar{b}} - \h \delta Y_{b \bar{b}} -
Y_2)}\,e^{\omega(\nu')\,(Y_2 - 0)} \notag\\
&& \times \,\,\,\,g(M_{W},-\nu_1)\,g(M_{W},\nu'_1) \,
g(M_{b\bar{b}},-\nu_2)\,g(M_{b\bar{b}},\nu'_2)\,  g\Lb \vec{k},
\vec{q}=0; \nu'  \Rb
 \nonumber
\eea

The  BFKL eigenvalues  $\omega(\nu) $ and $ \omega(\nu')$,  (see
\eq{OM})  increase as $\omega(\nu)  \to- \bas/(1/2 \pm i \nu)$, and
therefore, the integral over $\nu$ takes  the following form

\beq \label{FED31} \int^{+\infty - i (\epsilon-1/2)}_{- \infty - i (
\epsilon-1/2)}\,\,\frac{ d \nu}{2 \pi} \exp\Lb -\frac{\bas}{1/2 \pm
i \nu}\,( Y - Y_1)\Rb \to -\,\bas\,\,\int^{\infty }_{- \infty
}\,\,\frac{ d t }{2 \pi i}\,\,\frac{e^{t (Y - Y_1)}}{t^2} \,\,= \bas
(Y - Y_1) \eeq

where $t = \bas/(1/2 \pm i \nu)$ . The typical $\nu$ in this
integral, is about $\bas (Y-Y_1)$, while we expect $\nu_1 \approx
\nu_2 \approx 1/\bas Y$. Therefore, for region (i) \,\,\,\,
 $|i \nu + 1/2|$ cannot be less than $ \nu_1 + \nu_2$, which is in
 clear contradiction with our initial assumption.
 It  means, that the first kinematic region does not contribute to the integral.

\begin{boldmath}
\subsubsection{Region (II):
 \,\,\,\,  $|i \nu + 1/2| \gg \nu_1 + \nu_2$ ($ |i \nu' +
1/2| \gg \nu'_1 + \nu'_2$)}

\end{boldmath}
~

In the second kinematic region,  $|i \nu + 1/2| \gg  \nu_1 + \nu_2$
($ |i \nu '+ 1/2|  \gg  \nu_1' + \nu_2'$) while $\nu_1 \approx \nu_2
\approx \nu'_1 \approx \nu'_2 \approx 1/\as Y$. In this region,
\eq{GG} reads

 \bea \label{FED4}
&&\int\,d^2 q'\, \Gamma_{3\p} \Lb q=0,q'| \nu,\nu_1,\nu_2
\Rb\,\,\Gamma_{3\p} \Lb q=0,q'|
-\nu',-\nu'_1,-\nu'_2 \Rb \,\,\\
&&\,\,\,\,\,\,\,\,\,\,\,\,\,\,\,\,\longrightarrow\,\,\NA^2 \frac{2
\pi^9\,2^{2i\nu-2i\nu'} C_1(\nu)C_1(-\nu')}{( 1/2 + i \nu)^3(1/2 -
i\nu')^{3}}\,\delta ( \nu - \nu') \notag \eea

Taking into account \eq{FED4}, the integration over $\nu'$ is
eliminated by the Dirac delta function $\delta\Lb\,\nu-\nu'\Rb\,$,
which appears in \eq{FED4}, such that inserting this into \eq{FED1}
gives the following expression

\bea &&A_{\mbox{reg.} (ii)}\Lb \p
\to 2\p  \to \p;\fig{2psen} \Rb\,\, = \label{FED5}\\
&&\,\Sigma\Lb k,M_W,M_{b \bar{b}},k_0 \Rb\,\pi^8\intf \! \,
\,\frac{d \nu}{(1/2 + i
\nu)^3(1/2-i\nu)^3}\Lb\frac{k_0^2}{k^2}\Rb^{i\nu}\,
\,\prod^2_{i=1}\intf\,\nu^2_i\lambda\Lb\nu_i\Rb\,\, d
\nu_i\,\intf\,\,\nu'^2_i\lambda\Lb\nu_i'\Rb \, d \nu'_i \,\, \, \,\,
 \nonumber  \\
&&\times \,\, \,\int^{Y}_{Y_W + \h \delta
Y_W}\!\!\!\!\!\!\!dY_1\int^{Y_{b \bar{b}} - \h \delta Y_{b
\bar{b}}}_{0}\!\!\!\!\!\!\!dY_2\,\,\,e^{\omega(\nu)\,(Y -
Y_1+Y_2)}\,\,
  \, e^{\omega(\nu_1)\,( Y_1 - Y_W - \h \delta Y_W)} \,\, e^{\omega(\nu_2)\,( Y_1 - Y_{b \bar{b}} - \h \delta Y_{b \bar{b}})} \nonumber \\
&& \times \,\,\, e^{\omega(\nu'_1)\,(Y_W - \h \delta Y_W - Y_2)}\,
e^{\omega(\nu'_2)\,(Y_{b \bar{b}} - \h \delta Y_{b \bar{b}} -Y_2)}\,\,\,\,\,g(M_{W},-\nu_1)\,g(M_{W},\nu'_1) \, g(M_{b\bar{b}},-\nu_2)\,g(M_{b\bar{b}},\nu'_2) \nonumber \\
\eea

If the following definitions are made, namely

  \bea
Y_{\nu_1}\,\,\,&=&\,\,\,  Y_1 - Y_W - \h \delta
Y_W\,;\,\,\,L_{\nu_1} \,\,=\,\,\frac{M^2_W}{k^2_0}\,;\,\,\,
\,\,\,\,\,\,\,
\,\,\,\,\,\,\,\,\,\,\,\,\,\,\,Y_{\nu_2}\,\,\,=\,\,\,Y_1 - Y_{b \bar{b}} - \h \delta Y_{b \bar{b}}\,; \,\,\,L_{\nu_2} \,\,=\,\,\frac{M^2_{b \bar{b}}}{k^2_0}\,;   \notag\\
Y_{\nu'_1}\,\,\,&=&\,\,\, Y_W - \h \delta Y_W - Y_2;
\,\,\,L_{\nu'_1} \,\,=\,\,\frac{M^2_W}{k_0^2}\,;\,\,\,
\,\,\,\,\,\,\,\,\,\,\,\,\,\,\,\,\,\,\,\,\,\,\,
 Y_{\nu'_2}\,\,\,=\,\,\, Y_{b \bar{b}} - \h \delta Y_{b \bar{b}} - Y_2\,;\,\,\,L_{\nu'_2} \,\,=\,\,\frac{M^2_{b
 \bar{b}}}{k_0^2}\,;\,\,\,\notag
\label{Ynu} \eea

Then \eq{FED5} can be written in the following, more compact form,

\bea
&& A_{\mbox{reg.} (ii)}\Lb \p \to 2\p  \to \p;\fig{2psen} \Rb
=\notag\\
&&\,\,\Lb16\pi^4\Rb^4\,\Sigma\Lb k,M_W,M_{b \bar{b}},k_0
\Rb\int^{Y}_{Y_W + \h \delta
Y_W}\!\!\!\!\!\!\!dY_1\int^{Y_{b\bar{b}} - \h \delta Y_{b
\bar{b}}}_{0}\!\!\!\!\!\!\!dY_2\,I_\nu\,I_{\nu_1}\,I_{\nu_2}\,I_{\nu_1}'\,I_{\nu_2}'\label{FED5'}
\eea
  \bea\mbox{where}\,\,\,\,\, I_\nu(Y - Y_1 - Y_2)\,\,&=&\,\,\int^{\infty - i (
\epsilon-1/2)}_{- \infty - i ( \epsilon-1/2)}\! d \nu\,\frac{e^{
\frac{-\bas}{1/2 + i \nu}\,( Y - Y_1+Y_2)}}{(i \nu +
1/2)^3(1/2-i\nu)^3}\Lb\frac{k_0^2}{k^2}\Rb^{i\nu}\,\label{I1} \eea

\bea \mbox{and where}\,\,\,\,\,\,\,\,\,
 I_{\nu_i\to\,0}(\Lb Y_{\nu_i},L_{\nu_i}\Rb\,\,\,\,\,\,
\,\,&=&\,\,\,\,\Itn_i\,\nu_i\,e^{\omega(\nu_i)\, Y_{\nu_i}}\,\Lb
L_{\nu_i}\Rb^{ i\,\nu_i}\,\,\,\label{Ii2}\\
\notag\\
\mbox{comparison with
\eq{1P2a}}\,\,\,\,\,\,\,\,\Rightarrow\,\,\,\,\,\,\,
I_{\nu_i\to\,0}(\Lb
Y_{\nu_i},L_{\nu_i}\Rb&\xrightarrow{\nu_i\to\,0}&\,\,\,
\,\,\,\,\,\frac{d}{d\ln\,L_{\nu_1}} \,P\Lb L_{\nu_i}; Y_{\nu_i} \Rb\,\label{Ii}\\
\notag\\
 &\xrightarrow{Y_{\nu_i}\to\,0}&\,\,\,\,\frac{d}{d \ln L_{\nu_i}}\,\delta\Lb
\,\ln  L_{\nu_i}\,\Rb \eea

\vspace{0.2cm}

In \eq{FED5'}, we evaluate first the integrals over $Y_1$ and $Y_2$,
which take the following form, where in region (ii), $\nu_1$,
$\nu_2$, $\nu_1'$ and $\nu_2'$ are considered small.

\bea \label{YI} &&\int^{Y}_{Y_W + \h \delta
Y_W}\!\!\!\!\!\!\!dY_1\int^{Y_{b \bar{b}} - \h \delta Y_{b
\bar{b}}}_{0}\!\!\!\!\!\!\!dY_2\,I_\nu\,I_{\nu_1}\,I_{\nu_2}\,I_{\nu_1}'\,I_{\nu_2}'
\,\,=\intf\!\frac{d\nu}{(i \nu + 1/2)^3(i \nu-
1/2)^3}\Lb\frac{k_0^2}{k^2}\Rb^{i\nu}
\notag\\&\,&\times\,\,\,\intf\!\nu_1\,L_{\nu_1}^{-i\nu_1}d\nu_1\intf\!\frac{\nu_2\,L_{\nu_2}^{-i\nu_2}d\nu_2}{\Lb
\omega(\nu_1) + \omega(\nu_2) - \omega(\nu) \Rb\,}
\intf\!\nu_1'L_{\nu_1'}^{-i\nu_1'}d\nu_1'\intf\!\frac{\nu_2'\,L_{\nu_2'}^{-i\nu_2'}d\nu_2'}{\Lb
 \omega(\nu_1') +   \omega(\nu'_2)  -  \omega(\nu) \Rb} \,\, \notag \\
&&\times\,\,\,\,\,\,
 e^{\omega(\nu_1)\,( Y - Y_W - \h \delta Y_W)}\,
 e^{\omega(\nu_2)\,( Y - Y_{b \bar{b}} - \h \delta Y_{b \bar{b}})}\,\Lb\,1\,\,- \,\,
 \,e^{(\omega(\nu)\,- \,\omega(\nu_1) - \omega(\nu_2))( Y - Y_W - \h \delta Y_W)}\,\Rb \,\nonumber \\
 &&  \times\,\,\,\,\,\,
e^{\omega(\nu'_1)\,(  Y_{W} - \h \delta Y_{W})}\,
e^{\omega(\nu'_2)\,(Y_{b \bar{b}} - \h \delta Y_{b \bar{b}})}\,\
\Lb\,1\,\,-\,\, \, e^{( \omega(\nu)\,-\,\omega(\nu'_1) -
\omega(\nu'_2))( Y_{b \bar{b}} - \h \delta Y_{b \bar{b}})} \Rb \eea

In the limit that $i \nu \pm1/2 \to 0$, then
$\omega(\nu)\,\longrightarrow\,-\bas/(i\,\nu \pm 1/2)$ , and in our
kinematic region, \eq{YI} reduces to the following expression

 \bea
\label{YI1} &&\int^{Y}_{Y_W + \h \delta
Y_W}\!\!\!\!\!\!\!dY_1\int^{Y_{b \bar{b}} - \h \delta Y_{b
\bar{b}}}_{0}\!\!\!\!\!\!\!dY_2\,I_\nu\,I_{\nu_1}\,I_{\nu_2}\,I_{\nu_1}'\,I_{\nu_2}'
\,\,= \,\,
\frac{ 1}{\bas^2}\,\intf\!\frac{d\nu}{(i\,\nu \pm 1/2)(i \,\nu \mp 1/2)^3} \Lb\frac{k_0^2}{k^2}\Rb^{i\nu}\notag\\
&&\times\,\,\,\,\intf\!\nu_1L_{\nu_1}^{-i\nu_1}d\nu_1\intf\nu_2L_{\nu_2}^{-i\nu_2}d\nu_2\,
 e^{\omega(\nu_1)\,( Y - Y_W - \h \delta Y_W)}\,
 e^{\omega(\nu_2)\,( Y - Y_{b \bar{b}} - \h \delta Y_{b \bar{b}})}\,\nonumber\\
&& \times \,\,\,\Lb\,1\,\,- \,\,
 \,e^{(\omega(\nu)\,- \,\omega(\nu_1) - \omega(\nu_2))( Y - Y_W - \h \delta Y_W)}\,\Rb \,\nonumber \\
 &&  \times\,\,\intf\nu_1'L_{\nu_1'}^{-i\nu_1'}d\nu_1'\intf\nu_2'L_{\nu_2'}^{-i\nu_2'}d\nu_2'\,
e^{\omega(\nu'_1)\,(  Y_{b \bar{b}} - \h \delta Y_{b \bar{b}})}\,\,
e^{\omega(\nu'_2)\,(Y_{b \bar{b}} - \h \delta Y_{b \bar{b}})}\,\, \notag \\
&&\times\,\,\,\, \Lb\,1\,\,-\,\, \, e^{(
\omega(\nu)\,-\,\omega(\nu'_1) - \omega(\nu'_2))( Y_{b \bar{b}} - \h
\delta Y_{b \bar{b}})} \Rb \eea

In \eq{YI1},  the integration over $\nu$ can be taken analytically,
since

 \bea &&\int^{\infty - i ( \epsilon-1/2)}_{- \infty - i (
\epsilon-1/2)}\,\,\frac{ d \nu}{2 \pi\,i}\,\frac{1}{(i \nu \pm
1/2)\,\,(1/2\mp i\nu)^3}\,\,\,=\,\,1\,; \label{INU1}\\
&& \int^{\infty - i ( \epsilon-1/2)}_{- \infty - i (
\epsilon-1/2)}\,\,\frac{ d \nu}{2 \pi\,i}\,\,\frac{e^{-
\frac{\bas}{1/2\pm  i \nu}\, Y_{\nu}}}{(i \nu \pm 1/2)\,\,(1/2\mp
i\nu)^3}\Lb\frac{k_0^2}{k^2}\Rb^{i\nu} \,\,\,
\xrightarrow{k^2_0=k^2}\,\,\,\frac{1}{\bas}\,\int^{\infty }_{-\infty
}\,\,\frac{ d t }{2 \pi \,t}\,\,\,e^{-t Y_{\nu} }\notag\\ \,\,
&&\approx \,\, \Theta(\Lb Y - Y_1+Y_2 \Rb\,+\,\label{INU1'}\\
&&\,+\,\mbox{terms suppressed by powers of $\bas$ relative to the
first term (see \eq{INU1})} \label{FED6}\\
&&\rightarrow\,\,\frac{1}{2\pi}\left\{\sqrt{\frac{
\pi}{f''_{\mbox{sp}}}}\,\,\exp\Lb 2 \sqrt{\bas
\,Y_{\nu}\,\ln(k^2/k^2_0)}\Rb\right\} \Lb
1-\frac{\bas}{t_{\mbox{sp}}}\Rb^{-3} \label{FED61} \eea

 where $t=-
\bas/(1/2 \pm i \nu)$, and $\Theta(x) = 1$ if $x > 0$ while $
\Theta(x)=0$ for $x<0$. The last integral  is taken using the
steepest decent method, and  $t_{\mbox{sp}}=\sqrt{\frac{\bas
\,\ln(k^2/k^2_0)}{Y_i}}$ is the saddle point of the integral in
\eq{FED61}. $f''_{SP} \,=\,t_{SP}\,Y_i$. However, this method works
only in the region $Y_i\,t_{SP}\,>\,1$. For $Y_i\,t_{SP}\,< \,1$, we
can use \eq{FED6}. Using \eq{FED6} for the calculation, we find that
the largest contribution stems from the first term in both
parentheses in the integrand of \eq{YI1}. Hence, the first terms in
both parentheses survive, leading to the following expression

 \bea \label{YI2}
&&\int^{Y}_{Y_W + \h \delta Y_W}\!\!\!\!\!\!\!dY_1\int^{Y_{b
\bar{b}} - \h \delta Y_{b
\bar{b}}}_{0}\!\!\!\!\!\!\!dY_2\,\,\,I_\nu\,I_{\nu_1}\,I_{\nu_2}\,I_{\nu_1}'\,I_{\nu_2}'
\,\,= \notag\\ &&\,
\intf\nu_1L_{\nu_1}^{-i\nu_1}d\nu_1\intf\nu_2L_{\nu_2}^{-i\nu_2}d\nu_2\intf\nu_1'L_{\nu_1'}^{-i\nu_1'}d\nu_1'\intf\nu_2'L_{\nu_2'}^{-i\nu_2'}d\nu_2'\,\times\,\,\,
 \notag\\
&&\,\,\,\times\,\,\frac{1}{\bas^2}
 e^{\omega(\nu_1)\,( Y - Y_W - \h \delta Y_W)}\,
 e^{\omega(\nu_2)\,( Y - Y_{b \bar{b}} - \h \delta Y_{b \bar{b}})}\,\,\,
e^{\omega(\nu'_1)\,(  Y_{W} - \h \delta Y_{W})}\,\,
e^{\omega(\nu'_2)\,(Y_{b \bar{b}} - \h \delta Y_{b \bar{b}})}\,\,
\eea

The integrals over $\nu_i$ and $\nu'_i$, lead to the Pomeron
amplitudes given by \eq{1P2a}, and the final answer for this
kinematic region is given by the following expression.

 \bea \label{FED7}
&&A \Lb \p \to 2\p \to \p;\fig{2psen} \Rb_{\mbox{reg II}}\,\,\,=
\frac{ \Lb 2\pi\Rb^{16}}{\bas^2}\Sigma\Lb
k^2,M_W,M_{b\bar{b}},k_0\Rb\, \nu^{SP}_1\,
 \nu^{SP}_2\,\nu'^{SP}_1\,\nu'^{SP}_2\times\,\\
&&P\Lb k,M_{b , \bar{b}}| Y - Y_{b \bar{b}} - \h \delta Y_{b
\bar{b}}\Rb  P\Lb k_0,M_{b , \bar{b}}|  Y_{b \bar{b}} - \h \delta
Y_{b \bar{b}} \Rb \,\,\, P\Lb k,M_{W}| Y - Y_{W} - \h \delta
Y_{W}\Rb P\Lb k_0,M_{W}|  Y_{W} - \h \delta Y_{W}\Rb \nonumber \eea

 where $\nu^{SP}_i$ is the value of the saddle point in the integration over $\nu_i$, in the steepest decent method, and
$$\nu^{SP}_1\,
 \nu^{SP}_2\,\nu'^{SP}_1\,\nu'^{SP}_2 \,\,=\,\,\frac{\pi^2 \ln(M^2_W/k^2)\, \ln(M^2_W/k^2_0)\,\ln(M^2_{b
\bar{b}}/k^2)\,\ln(M^2_{b \bar{b}}/k^2_0)}{ \Lb Y - Y_{b \bar{b}} -
\h \delta Y_{b \bar{b}}\Rb \,\Lb Y - Y_{W} - \h \delta Y_{W}
\Rb\,\Lb  Y_{b \bar{b}} - \h \delta Y_{b \bar{b}} \Rb\,\Lb Y_{W} -
\h \delta Y_{W}\Rb} $$

\begin{boldmath}
\subsubsection{Region (III):\,\,\,\, $|i \nu + 1/2| \approx \nu_1 + \nu_2$ ($ |i
\nu'+ 1/2| \approx\nu'_1 + \nu'_2$)}
\end{boldmath}
As it has been pointed out in Ref.\cite{LMP}, the fastest increase
with energy stems from the third  region of integration over $\nu$,
$\nu_1$ and $\nu_2$, which is specified by

 \bea \label{DOMRE}
&\nu_1\,\,\to \,\,\nu_2\,;\,\,\,\,\,1/2 + i \nu - i \nu_1 - i
\nu_2\,\,=\,\,0\,;
\,\,\,\,\,\,\,\,\omega( 2\nu^0_1 + \frac{1}{2}\,i)\,\,\,=\,\,\,2\,\omega(\nu^0_1)& \nonumber \\
&\nu'_1\,\,\to \,\,\nu'_2\,;\,\,\,\,\,1/2 + i \nu' - i \nu'_1 - i
\nu'_2\,\,=\,\,0\,; \,\,\,\,\,\,\,\,\omega( 2\nu'^0_1 +
\frac{1}{2}\,i)\,\,\,=\,\,\,2\,\omega(\nu'^0_1)& \eea

It turns out, that $2 \omega(\nu^0_1) \,>\,2 \omega(0)$, and
therefore the contribution from the kinematic region specified by
\eq{DOMRE}, leads to a faster growth than in the case of the
exchange of two non interacting Pomerons. It means, that this region
can lead to a larger value for the cross section  for the two parton
shower production at the LHC energy, in comparison with the Tevatron
energy, (where such a cross section has been measured in ref.
\cite{DPXS}).  We can  integrate  over $Y_1$ and $Y_2$ in \eq{FED1},
to give the result

 \bea &&A\Lb \p \to 2\p \to \p;\fig{2psen} \Rb =\Itn
\,\,\nu^2\,\, \,\prod^2_{i=1}\Itn_i\nu^2_i\,\,\lambda\Lb\nu_i\Rb
\,\Itnpr_{i}\nu'^2_i\,\lambda(\nu'_i)\, \Itnpr
\, \, \nu'^2\,  \label{FED8} \\
&&\times  \,g_P\Lb \vec{k}, \vec{q}=0; -\nu  \Rb \int\,d^2 \,q' \,\,\Gamma_{3\p}\Lb \vec{q}=0,\vec{q}\,'|\nu,\nu_1, \nu_2 \Rb\,\, \Gamma_{3\p}\Lb \vec{q}=0,\vec{q}\,'|-\nu',-\nu'_1,- \nu'_2 \Rb\,\,g_P\Lb \vec{k}_0, \vec{q}=0 ;\nu'  \Rb\notag\\
&&\times\,\sigma\Lb\,M_w,-\nu_1,\nu_1'\Rb\sigma\Lb\,M_{b\bar{b}},-\nu_2,\nu_2'\Rb\,\Lb
\omega(\nu_1) +   \omega(\nu_2)  -  \omega(\nu) \Rb^{-1}\,\,\Lb
  \omega(\nu')-\omega(\nu'_1) - \omega(\nu'_2) \Rb^{-1} \nonumber\\
 && \times \Lb e^{\omega(\nu_1)\,( Y - Y_W - \h \delta Y_W)}\,
 e^{\omega(\nu_2)\,( Y - Y_{b \bar{b}} - \h \delta Y_{b \bar{b}})}\,\,-\,\,
 \,e^{\omega(\nu)\,( Y - Y_W - \h \delta Y_W)}\,
  e^{(\omega(\nu_2)\,(  Y_W +  \h \delta Y_W - Y_{b \bar{b}} - \h \delta Y_{b \bar{b}})}\Rb \,\nonumber \\
 &&  \times \Lb
e^{(\omega(\nu'_1)\,(  Y_W -  \h \delta Y_W - Y_{b \bar{b}} + \h
\delta Y_{b \bar{b}})}\, e^{\omega(\nu')\,(  Y_{b \bar{b}} - \h
\delta Y_{b \bar{b}})}\,\,-\,\, e^{(\omega(\nu'_1)\,(  Y_W -  \h
\delta Y_W )}\, e^{\omega(\nu'_2)\,( Y_{b \bar{b}} - \h \delta Y_{b
\bar{b}})}\Rb \nonumber \eea

Using \eq{sigma} and \eq{GG}, and the expression given in the
appendix (see \eq{Ge0}), we obtain

 \bea
&&A\Lb \p \to 2\p \to \p;\fig{2psen} \Rb =\,\pi^{8}\,\Sigma\Lb
k,M_W,M_{b \bar{b}},k\Rb\,\intf \! d \nu\,\intf\!d \nu'\,
\,\prod^2_{i=1}\,\,\intf\! \nu^2_i\,\lambda\Lb\nu_i\Rb\,d \nu_i\,\intf\!\nu'^2_i\lambda\Lb\nu_i'\Rb\,\,d \nu'_i\nonumber\\
&&\times\,\,\,\Lb 16k^2_0\Rb^{i \nu'}\Lb 16k^2\Rb^{-i \nu}  \Lb
16M^2_W\Rb^{ i \nu_1 '-i \nu_1} \Lb 16M^2_{b \bar{b}}\Rb^{ i \nu_2
'- i \nu_2}\,\,
\, \, \, \delta( \nu - \nu_1 - \nu_2 - \nu' + \nu'_1 + \nu'_2) \nonumber \\
&&\times  \,\,\frac{\Lb\, e^{\omega(\nu_1)\,( Y - Y_W - \h \delta
Y_W)}\,
 e^{\omega(\nu_2)\,( Y - Y_{b \bar{b}} - \h \delta Y_{b \bar{b}})}\,\,-\,\,
 \,e^{\omega(\nu)\,( Y - Y_W - \h \delta Y_W)}\,
  e^{\omega(\nu_2)\,(  Y_W +  \h \delta Y_W - Y_{b \bar{b}} - \h \delta Y_{b \bar{b}})\, }\Rb}{\Lb 1/2 - i \nu +i\nu_1 + i \nu_2\Rb^3\,\Lb  \omega(\nu_1) +   \omega(\nu_2)  -  \omega(\nu) \Rb\,} \nonumber\\
 &&  \times \frac{\Lb
e^{\omega(\nu'_1)\,(  Y_W -  \h \delta Y_W - Y_{b \bar{b}} + \h
\delta Y_{b \bar{b}})}\, e^{\omega(\nu')\,(  Y_{b \bar{b}} - \h
\delta Y_{b \bar{b}})}\,\,-\,\, e^{\omega(\nu'_1)\,(  Y_W -  \h
\delta Y_W )}\,
e^{\omega(\nu'_2)\,( Y_{b \bar{b}} - \h \delta Y_{b \bar{b}})}\Rb}{ \Lb 1/2 +i \nu '-i\nu_1' - i \nu_2'\Rb^3    \Lb\,\omega(\nu')-\omega(\nu'_1) -  \omega(\nu'_2) \Rb} \nonumber\\
&&\times\,C_1\Lb\,-\nu_1\Rb\,C_1\Lb\,\nu_1'\Rb\,C_1\Lb\,-\nu_2\Rb\,C_1\Lb\,\nu_2'\Rb\label{FED9}
\eea

In order to evaluate the integrations over $\nu_1$, $\nu_2$,
$\nu_1'$ and $\nu_2'$ on the RHS of \eq{FED9}, it will be useful to
make the following change of variables. Let
$\nu_{(12)}\,=\h\,\Lb\,\nu_1+\nu_2\Rb\,$, and
$\nu_{\,[12]}\,=\,\h\Lb\,\nu_1-\nu_2\Rb\,$, and likewise for
$\nu_{(12)}'\,=\h\,\Lb\,\nu_1+\nu_2\Rb\,$, and
$\nu_{\,[12]}'\,=\,\h\Lb\,\nu_1'-\nu_2'\Rb\,$. In new variables $
\,\prod^2_{i=1}\,\nu^2_i\,\lambda\Lb\nu_i\Rb\,d
\nu_i\,\nu'^2_i\lambda\Lb\nu_i'\Rb\,\,d \nu'_i$ has the form $16
\Lb\nu_{(12)}^2-\nu_{[\,12]}^2\Rb\,\,\,
\Lb\nu_{(12)}'^2-\nu_{[\,12]}'^2\Rb\,\,\,d \nu_{(12)} d
\nu_{\,[12]}\,d \nu'_{(12)} d \nu'_{\,[12]}$. We expect that
$\nu_{(12)}$ and $\nu'_{(12)}$ will be close to $\nu^0_1$ while both
$\nu_{\,[12]}$ and $\nu'_{\,[12]}$ will be much smaller. Indeed,
assuming this  we can expand $\omega(\nu_i) =
\omega(\nu_{(12)})\,+\,\omega'(\nu_{(12)})\,\nu_{\,[12]}\,\,-\,\,\omega"(\nu_{(12)})\,\nu^2_{\,[12]}/2$.
Using this expansion one can see that integrals over $\nu_{[\,12]}$
and $\nu_{[\,12]}$ have a Gaussian form and can be taken using the
steepest decent method. The saddle point value of these variables
are small and decreasing with energy, namely,
\beq \label{SADDLE3}
\nu^{SP}_{[\,12]}\,\,=\,\,\frac{\ln\Lb M^2_W\,M^2_{b
\bar{b}}/k^4_0\Rb}{2\,Y\,\,-\,\, Y_W  -\h \delta Y_W - Y_{b \bar{b}}
-\h \delta Y_{b \bar{b}}}\,;\,\,\,\,\,
\nu'^{SP}_{[\,12]}\,\,=\,\,\frac{\ln\Lb M^2_W\,M^2_{b
\bar{b}}/k^4_0\Rb}{\,\, Y_W  -\h \delta Y_W + Y_{b \bar{b}} -\h
\delta Y_{b \bar{b}}}\,; \eeq

Evaluating the integrals over $\nu_{(12)}$ and $\nu'_{(12)}$ by
closing the contour on the pole, and  by using the $\delta$
function, \eq{FED9} reduces to the following expression (considering
$\nu^{SP}_{[\,12]}$  as well as  $\nu'^{SP}_{[\,12]}$ are small:
$\nu^{SP}_{[\,12]})\,\ll\,\nu_{(12)}$ and
$\nu'^{SP}_{[\,12]})\,\ll\,\nu'_{(12)}$).

\bea &&A\Lb \p \to 2\p \to \p;\fig{2psen} \Rb =\,\pi^{8}\,
\,\Sigma\Lb k,M_W,M_{b \bar{b}},k_0\Rb\, \int \! \! d \nu\,\,d
\nu'\,\, \nu_1\nu_2\,\, d \nu_{12} \,\,\nu'_1\nu'_2\,\, d
\nu'_{12}\,\,\,\,
\, \, \,  \nonumber \\
&&\times  \, \Lb\frac{ k^2}{M^2_W\,M^2_{b \bar{b}}}\Rb^{ - i
\nu}\Lb\frac{ k^2_0}{M^2_W\,M^2_{b \bar{b}}}\Rb^{ i \nu'} \frac{
\omega''(0) \,( Y - Y_W - \h \delta Y_W)}{\Lb \omega(\nu_1) +
\omega(\nu_2)  -  \omega(\nu) \Rb\,\,\Lb
 \omega(\nu'_1) +   \omega(\nu'_2)  -  \omega(\nu') \Rb} \nonumber\\
 && \times\,\,\,\,
 e^{\omega(\nu_1)\,( Y - Y_W - \h \delta Y_W)}\,
 e^{\omega(\nu_2)\,( Y - Y_{b \bar{b}} - \h \delta Y_{b \bar{b}})}\,\Lb\,1\,\,- \,\,
 \,e^{(\omega(\nu)\,- \,\omega(\nu_1) - \omega(\nu_2))( Y - Y_W - \h \delta Y_W)}\,\Rb \,\nonumber \\
 &&  \times\,\,\,\,
e^{\omega(\nu'_1)\,(  Y_{W} - \h \delta Y_{W})}\,
e^{\omega(\nu'_2)\,(Y_{b \bar{b}} - \h \delta Y_{b \bar{b}})}\,\
\Lb\,1\,\,-\,\, \, e^{( \omega(\nu')\,-\,\omega(\nu'_1) -
\omega(\nu'_2))( Y_{b \bar{b}} - \h \delta Y_{b \bar{b}})} \Rb
\label{FED10} \eea

In this integral $\nu_{(12)}\,\,=\,\,\h(\nu - \h \,i)$ and
$\nu'_{(12)}\,\,=\,\,\h(\nu' - \h\, i)$ and $\nu_i = \h(\nu - \h
\,i) \pm  \nu_{\,[12]}\,\,\to\,\,\nu^0_1$ at high energy, as well as
$\nu'_i = \h(\nu' - \h \,i) \pm  \nu'_{\,[12]}\,\,\to\,\,\nu^0_1$.
 It should be
stressed, that in \eq{FED10}, all the $\nu$'s satisfy \eq{DOMRE}.
Expanding $\omega(\nu_i)$ around the point in the vicinity of
$\nu_1= \nu'_1 = \nu^0_1 $, and denoting $ i \nu_1 - i\nu^0_1 = f$
and $ i \nu_1 - i\nu^0_1 = f' $, we find that \eq{FED2} can be
written as follows

 \bea\label{EDF1} && \int^{ \epsilon + i
\infty}_{\epsilon -i \infty} \frac{d f}{2 \pi i}\,\, e^{
(\omega(\nu^0_1)\,+\,\tilde{d}\,f/2  )\,( 2 Y \,\,-\,\, Y_W - Y_{b
\bar{b}} - \h( \delta  Y_W  + \delta Y_{b \bar{b}} ))} \,\frac{1}{d
\,f}\,\,
\Lb 1 -  e^{- d\,f \,( Y -Y_W - \h \delta Y_W )} \Rb\, \nonumber \\
 &&
\int^{ \epsilon + i \infty}_{\epsilon -i \infty} \frac{d f'}{2 \pi
i}\,\, e^{ (\omega(\nu^0_1)\,+\,\tilde{d}\,f'/2 )\, ( Y_W + Y_{b
\bar{b}} - \h (\delta Y_M + \delta Y_{b \bar{b}})}\,\,\frac{1}{d
\,f'}\,\Lb 1 \,\,-\,\,e^{ - d f' \,( Y_{b \bar{b}} + \h
  \delta Y_{b \bar{b}}) } \Rb
\eea

 where $\tilde{d }\equiv 2 \omega"(\nu_1 = \nu^0_1)\,>\,0$ and $
d \equiv   \tilde{d} - \omega"(2 \nu^0_1 + i 1/2\, >\, \tilde{d}$.
In this integral, we can close the contour of integration over $f$
and $f'$, on the poles $f=0$ and $f'=0$, for the first term in each
set of parentheses. In all other terms in \eq{EDF1}, we can close
the contour on the right semi-plane, where we have no singularities,
and therefore these integrals are equal to zero. Finally, we have

\bea \label{FED11} &&A\Lb \p \to 2\p \to \p;\fig{2psen} \Rb\,\,\,\,
=
 \,\,\,\pi^{8}\,\Sigma\Lb k,M_W,M_{b \bar{b}},k_0 \Rb\\
&&\,\frac{ (\nu^0_1)^4 }{d^2}\int \! \!\!
 \frac{d \nu_{12}\,\,d \nu'_{12}}{\sqrt{k^2 k^2_0 M^2_W M^2_{b \bar{b}}}}
\, \, \,
  \, \Lb\frac{ k^2}{\sqrt{M^2_W\,M^2_{b \bar{b}}}}\Rb^{2 \nu^0_1}\Lb\frac{ k^2_0}{\sqrt{M^2_W\,M^2_{b \bar{b}}}}\Rb^{2 \nu^0_1}
 \omega''(0) \,( Y - Y_W - \h \delta Y_W)  \nonumber \\
 && \times
 e^{\omega(\nu^0_1 + \h \nu_{12})\,( Y - Y_W - \h \delta Y_W)}\,
 e^{\omega(\nu^0_1 - \h \nu_{12})\,( Y - Y_{b \bar{b}} - \h \delta Y_{b \bar{b}})}\, \,\,\,
e^{\omega(\nu^0_1 + \h \nu'_{12})\,(  Y_{W} - \h \delta Y_{W})}\,
e^{\omega(\nu^0_1 - \h \nu'_{12})\,(Y_{b \bar{b}} - \h \delta Y_{b
\bar{b}})}\ \nonumber \eea

(
 in \eq{FED11}, we hope that the factor $d$ will be not confused with the sign of the differential)
 Expanding   $\omega(\nu^0_1 + \h \nu_{12})$ and  $\omega(\nu^0_1 +
\h \nu'_{12})$ for the case of small $\nu_{12} $ ($\nu'_{12}$), and
integrating over
 $\nu_{12} $ and $\nu'_{12}$, we obtain

\bea \label{FED12} &&A\Lb \p \to 2\p \to \p;\fig{2psen} \Rb
\,\,\,\,=\,\,\,\,\,\pi^{9}\,\Sigma\Lb k,M_W,M_{b \bar{b}},k_0
\Rb\,\,\frac{ \,(\nu^0_1)^4}{d^2} \,  \, \, \,
  \, \Lb\frac{ k^2}{\sqrt{M^2_W\,M^2_{b \bar{b}}}}\Rb^{2 \nu^0_1}\Lb\frac{ k^2_0}{\sqrt{M^2_W\,M^2_{b \bar{b}}}}\Rb^{2 \nu^0_1}
\, \nonumber \\
&&\times\, \omega''(0) \,( Y - Y_W - \h \delta Y_W) \,\,\,
 \exp\Lb 2\,\,\omega(\nu^0_1 )\,( Y  - \h (\delta Y_W + \delta Y_{b \bar{b}})) \Rb\,\, \nonumber \\
 &&\times\,\frac{1}{\sqrt{2\,\omega"(0) ( 2 Y  - Y_W  -  Y_{b \bar{b}})}}\,
\,\,
 \exp \Lb - \frac{\ln^2\Lb M_W\,M_{b \bar{b}} /k^2\Rb}{2 \omega"(0) \Lb 2\,Y -  ( Y_W + Y_{b \bar{b}})\Rb} \Rb \,
 \nonumber \\
&&\times\,
  \frac{1}{\sqrt{2\,\omega"(0) ( Y_W +  Y_{b \bar{b}})}}
\exp \Lb - \frac{\ln^2\Lb M_W\,M_{b \bar{b}} /k^2_0\Rb}{2 \omega"(0)
\Lb Y_W + Y_{b \bar{b}}\Rb} \Rb \eea

\eq{FED12} can be rewritten in a simpler way, namely

\bea \label{FED13} &&A\Lb \p \to 2\p \to \p;\fig{2psen}
\Rb_{\mbox{reg. III}} \,\,\,\,=\,\,\,\,\,\pi^{9}\,\Sigma\Lb
k,M_W,M_{b \bar{b}},k_0 \Rb\,\,\frac{ \,(\nu^0_1)^4}{d^2}\,
  \, \Lb\frac{ k^2}{\sqrt{M^2_W\,M^2_{b \bar{b}}}}\Rb^{2 \nu^0_1}\Lb\frac{ k^2_0}{\sqrt{M^2_W\,M^2_{b \bar{b}}}}\Rb^{2 \nu^0_1}
\, \nonumber \\
&&\times\, \omega''(0) \,( Y - Y_W - \h \delta Y_W) \,\,\,
 \exp\Lb 2\,\,\Lb \omega(\nu^0_1 ) \,-\,\omega(0) \Rb\,\Lb Y  - \h (\delta Y_W + \delta Y_{b \bar{b}})) \Rb\,\,\Rb  \\
 &&\times\,P \Lb k, \sqrt{M_W\,M_{b \bar{b}}} |  2 Y  - Y_W  -  Y_{b \bar{b}}  - \h( \delta Y_W  -  \delta Y_{b \bar{b}})\Rb\,\,
P \Lb k_0, \sqrt{M_W\,M_{b \bar{b}}} |  Y_W  +  Y_{b \bar{b}} - \h(
\delta Y_W  -  \delta Y_{b \bar{b}})\Rb \notag\eea

\begin{boldmath}
\subsubsection{Region (IV): \,\,\,$ \nu \ll 1$, $\nu_1  \ll  1$ and  $\nu_2 \ll 1$ ($ \nu' \ll 1$, $\nu'_1  \ll  1$ and  $\nu'_2 \ll 1$ )}
\end{boldmath}
In our previous calculations, we focused our efforts on the singular
part of $\Gamma_{3\p} \Lb q=0,q'| \nu,\nu_1,\nu_2 \Rb
 \equiv \NA\int\,d^2 k\,\,\,g\Lb k,q=0,\nu \Rb \,\,g\Lb  \kv + \frac{1}{2} \qv\,', \qv\,' , \nu_1\Rb \,\,
g\Lb  \kv + \frac{1}{2}  \qv\,',- \qv\,' ,\nu_2\Rb$ (see \eq{G3P}).
However, the integral in the region where all the $\nu$'s are small,
also can lead to a contribution which  increases at large values of
$Y$, as $\exp\Lb \omega(0) \Lb 2 Y - \h \delta Y_W - \h \delta Y_{b
\bar{b}} \Rb \Rb$. Indeed, in this kinematic region we can integrate
\eq{FED31} over all  $ \nu_i$ ($\nu'_i$), using the expansion of
\eq{OMe}. As an example, the $\nu_1$ integration in \eq{FED311}
takes the following form, using the expansion of \eq{OMe}.

\bea \Itn_1\,\nu_1\lambda\Lb \nu_1\Rb\,e^{\omega\Lb 0\Rb\Lb
Y_1-Y_W-\h\delta\,Y_W\Rb+\h\nu_1^2\omega''\Lb0\Rb\Lb
Y_1-Y_W-\h\delta\,Y_W\Rb-i\nu_1\ln \Lb M_W^2/k^2\Rb} \label{eg}\eea

The exponential has a saddle point at the point
$\nu_{1}^{\mbox{\footnotesize{sp}}}\,=\,i\ln \Lb
M_W^2/k_0^2\Rb\,/\,\Lb\omega''\Lb 0\Rb\Lb
Y_1-Y_W-\h\delta\,Y_W\Rb\Rb$. The expression in the exponential
$f\Lb \nu_1\Rb$, at the saddle point takes the value\\ $f\Lb
\nu_1^{\mbox{\footnotesize{sp}}}\Rb\,=\,\h\ln^2\Lb
M_W^2/k^2\Rb\,/\,\Lb\omega''\Lb 0\Rb\Lb
Y_1-Y_W-\h\delta\,Y_W\Rb\Rb$. Hence, evaluating the integration in
\eq{eg}, using the method of steepest descents gives the following
result.

\bea &&16\,\nu_1^{\mbox{\footnotesize{sp}}}e^{\omega\Lb 0\Rb\Lb
Y_1-Y_W-\h\delta\,Y_W\Rb}\times\,e^{f\Lb\nu_1^{\mbox{\footnotesize{sp}}}\Rb}\Lb\frac{2\pi}{f''\Lb\nu_1^{\mbox{\footnotesize{sp}}}\Rb}\Rb^{1/2}\,=\,16\,\frac{i\ln
\Lb M_W^2/k^2_0\Rb\,}{\,\Lb\omega''\Lb
0\Rb\Lb Y_1-Y_W-\h\delta\,Y_W\Rb\Rb}\label{eg2}\\
&&\times\,e^{\omega\Lb 0\Rb\Lb
Y_1-Y_W-\h\delta\,Y_W\Rb}\times\,e^{\h\ln^2\Lb
M_W^2/k^2_0\Rb\Lb\omega''\Lb 0\Rb\Lb
Y_1-Y_W-\h\delta\,Y_W\Rb\Rb^{-1}}
\,\times\,\Lb\frac{2\pi}{\omega''\Lb 0\Rb\Lb
Y_1-Y_W-\h\delta\,Y_W\Rb}\Rb^{1/2}\notag
 \eea

 where $f''\Lb
\nu_1^{\mbox{\footnotesize{sp}}}\Rb\,=\,\omega''\Lb 0\Rb\Lb
Y_1-Y_W-\h\delta\,Y_W\Rb$. The factor of 16 in front in \eq{eg2},
originates from the value which $\lambda\Lb \nu_1\Rb$ appearing in
\eq{eg} takes at small $\nu_1$, (see \eq{BFKLLA} ). The integration
over the $\nu_1$, $\nu_1'$ and $\nu_2'$, are evaluated in a similar
way, using the method of steepest descents. Hence, the result after
integration of the RHS of \eq{FED31}, can be written as follows

  \bea
&&A\Lb \p \to 2\p \to \p;\fig{2psen} \Rb_{\mbox{\footnotesize{reg\,
(iv)}}}\,\,=\,\,\,\,\Lb 16\,\pi^4\Rb^4\Sigma\Lb k,M_W,M_{b
\bar{b}},k_0
\Rb \label{R41}\\
&& \times\,\,\, \int^{Y}_{Y_W + \h \delta
Y_W}\!\!\!\!\!\!\!dY_1\int^{Y_{b \bar{b}} - \h \delta Y_{b
\bar{b}}}_{0}\!\!\!\!\!\!\!dY_2\,\, \,e^{\omega(0)\,(Y -
Y_1+Y_2)}\,\,
  \, e^{2\omega(0)\,( Y_1 -Y_2 -\h \delta Y_W-\h\delta\,Y_{b\bar{b}})}\notag \\
&& \times\,\,\ , \Lb\frac{2\pi}{\omega''(0) \Lb Y -
Y_1\Rb}\Rb^{1/2}\,\,\Lb\frac{2\pi}{\omega''(0) \Lb Y_2 -
0\Rb}\Rb^{1/2}\,\ln^2\Lb M^2_W/k^2_0\Rb\,\ln^2 \Lb M^2_{b
\bar{b}}/k^2_0\Rb
\nonumber \\
&&\times\,\,\,2\pi\left\{\omega''(0) \Lb Y_1 - Y_W - \h \delta
Y_W\Rb\,\omega''\Lb0\Rb\Lb Y_{W} - \h \delta Y_{W} - Y_2 \Rb
\right\}^{-3/2}\nonumber\\
 && \times\,\,\,2\pi\left\{\omega''(0) \Lb Y_1
- Y_{b \bar{b}} - \h \delta Y_{b \bar{b}}\Rb \,\omega''\Lb 0\Rb\Lb
Y_{b \bar{b}} - \h \delta Y_{b
\bar{b}} -Y_2 \Rb\,\right\}^{-3/2}\notag\\
&&\times\,\,\,\exp\Lb-\h\frac{\ln^2\Lb M_W^2/k^2_0\Rb}{\omega''\Lb
0\Rb\Lb Y_1-Y_W-\h\delta\,Y_W\Rb}\Rb\,\exp\Lb-\h\frac{\ln^2\Lb
M_W^2/k^2_0\Rb}{\omega''\Lb 0\Rb\Lb
Y_W-\h\delta\,Y_W-Y_2\Rb}\Rb\nonumber\\
&&\times\,\,\,\exp\Lb -\h\frac{\ln^2\Lb
M_{b\bar{b}}^2/k^2_0\Rb}{\omega''\Lb 0\Rb \Lb
Y_1-Y_{b\bar{b}}-\h\delta\,Y_{b\bar{b}}\Rb}\Rb\,\exp\Lb-\h\frac{\ln^2\Lb
M_{b\bar{b}}^2/k^2_0\Rb}{\omega''\Lb 0\Rb\Lb
Y_{b\bar{b}}-\h\delta\,Y_{b\bar{b}}-Y_2\Rb}\Rb\nonumber
 \eea

The largest contribution to the integrals over $Y_1$ and $Y_2$,
stems from the region $Y - Y_1 \propto 1/\omega(0) $, and $Y_2 - 0
\propto 1/\omega(0)$. Since $\omega"(0)/\omega(0) \approx 1$, as far
as the QCD coupling is concerned, we can conclude from \eq{R41} that
 (using the \eq{1P2'} for the expression for $P\Lb
k,k_0\vert\,Y-Y'\Rb$, for the case of small $\nu$ )

\bea \label{R42} &&A \Lb \p \to 2\p \to \p;\eq{R41}
\Rb_{\mbox{reg. IV}}\,\,\,\propto\,\, \,(2 \pi)^{16}\,\Sigma\Lb k,M_W,M_{b \bar{b}},k_0 \Rb\\
&&\times \,\,\frac{\pi^2 \ln(M^2_W/k^2_0)\,
\ln(M^2_W/k^2_0)\,\ln(M^2_{b \bar{b}}/k^2_0)\,\ln(M^2_{b
\bar{b}}/k^2_0)}{ \Lb Y - Y_{b \bar{b}} - \h \delta Y_{b \bar{b}}\Rb
\,\Lb Y - Y_{W} - \h \delta Y_{W} \Rb\,\Lb  Y_{b \bar{b}} - \h
\delta Y_{b \bar{b}} \Rb\,\Lb Y_{W} - \h \delta Y_{W}\Rb} \,
P\Lb k,M_{b , \bar{b}}| Y - Y_{b \bar{b}} - \h \delta Y_{b \bar{b}}\Rb  \nonumber \\
&&\times\,\, P\Lb k_0,M_{b , \bar{b}}|  Y_{b \bar{b}} - \h \delta
Y_{b \bar{b}} \Rb \,\,\, P\Lb k,M_{W}| Y - Y_{W} - \h \delta
Y_{W}\Rb \,\,\,P\Lb k_0,M_{W}|  Y_{W} - \h \delta Y_{W}\Rb \nonumber
 \eea

 Comparing this contribution with \eq{FED7}, one can see that

\beq \label{R43}
 \frac{A \Lb \p \to 2\p \to \p;\eq{R42} \Rb_{\mbox{reg IV}}}{ A \Lb
\p \to 2\p \to \p;\eq{FED7} \Rb_{\mbox{reg
II}}}\,\,\,\propto\,\,\,\bas^2\,\,\ll\,\,1 \eeq

Therefore, this kinematic region gives a much smaller contribution,
than the kinematic region (II).

\numberwithin{equation}{section}
\section{Estimates of the background}

Comparing \eq{FED7} with \eq{FED13}, one can see that the first
contribution is suppressed. Indeed, the ratio of these two
contributions for $Y_W = Y_{b \bar{b}} =0$ in the c.m.f.,   is equal
to

\bea \label{EB} \frac{A\Lb \p \to 2\p \to \p;\eq{FED13}
\Rb{\mbox{reg III}}}{A\Lb \p \to 2\p  \to \p;\eq{FED7} \Rb{\mbox{reg
II}}} &=& \bas^2\omega"^2(0) \,( Y/2  - \h \delta Y_W)\,\,
(  Y/2  - \h \delta Y_{b \bar{b}}) \,\,\frac{(\nu^0_1)^4}{d^2\,\,\nu^{SP}_1 \,\nu^{SP}_2\,\nu'^{SP}_1\,\nu'^{SP}_2}  \notag \\
&\times &
 \exp\Lb 2\,\,\Lb \omega(\nu^0_1 ) \,-\,\omega(0) \Rb\,\Lb Y  - \h (\delta Y_W + \delta Y_{b \bar{b}})) \Rb\,\,\Rb
\,\,\gg\,\,1 \eea

 Indeed, the main factor that determines the energy
behavior,  is \\ $ \exp\Lb 2\Lb \omega(\nu^0_1 ) \,-\,\omega(0)
\Rb\,\Lb Y  - \h (\delta Y_W + \delta Y_{b \bar{b}})
\Rb\,\Rb\,\gg\,\,1$ which increases with energies. Other factors
include $\omega"^2(0)/d^2 \approx 1/4$,  and $   \bas\,( Y/2  - \h
\delta Y_W) \,\,\gg\,\,1$, as well as $\bas\, (  Y/2  - \h \delta
Y_{b \bar{b}})\,\,\gg\,\,1$. However , at high energy
$\nu^0_1/\nu^{SP}_1 \,\,\gg \,\,1$. This factor cannot change the
asymptotic behavior of this ratio, but it leads to a decrease in the
range of energy that we are interested in, namely from the Tevatron
to the LHC energy.

It turns out, that
 both at the Tevatron  and the LHC energies, the ratio of  \eq{EB}
is   smaller than unity , since $\nu^{SP}_1$ is equal to $0.37$ and
$0.19$, respectively,
 if we take $\as = \as\Lb M_H\,k_0\Rb$,
and $k_0 = 1\,GeV$. Such values of $\nu_{SP}$ lead to
$\nu^0_1/\nu^{SP}_1 = 0.21$ at the Tevatron, and
$\nu^0_1/\nu^{SP}_1 = 0.42$ at the LHC.
 The numerical solution of \eq{DOMRE}, gives the following relation (see Ref.
\cite{LMP})

\beq 2 ( \omega(\nu^0_1) - \omega(0))  \approx 0.25\,\bas\eeq

which leads to $ \exp\Lb 2\,\,\Lb \omega(\nu^0_1 ) \,-\,\omega(0)
\Rb\,\Lb Y  - \h (\delta Y_W + \delta Y_{b \bar{b}}))
\Rb\,\,\Rb\,\,\,\approx\,\,1.4 (\mbox{for the
Tevatron\,\,\,\,\,\,\,}$ and $1.86$ for the LHC. Collecting these,
estimate we obtain that the ratio of \eq{EB} to be equal to
$7\,\times\,10^{-4}$ for the Tevatron, and $1.4\,\times\,10^{-2}$
for the LHC. It means that we can neglect the contribution of
\eq{FED13},  and consider only the contribution given by \eq{FED7},
which describes the  production of the $W$ boson, and the $b
\bar{b}$-pair from two parton showers.

In this paper,  we obtain two different expression for the two
parton shower production, namely \eq{E2} for the non-enhanced
diagram, and \eq{FED7} for the enhanced diagram. From  a purely
theoretical point of view, it has been proven \cite{BART,BRN}, that
the only contribution stems from the enhanced diagram, which
reproduces the non-enhanced diagram in the low energy limit ($ Y -
y_1 \approx 1/\as, y_2 \approx 1/\as$), at least for the case of
dipole-dipole scattering. In this case, their contribution given by
\eq{FED7}, has been measured at the Tevatron \cite{DPXS}, and  it's
value at the LHC energy can be estimated using the following
expression

\bea
 \frac{ d^2 \sigma \Lb p + p \to W + [b \bar{b}] + X\Rb}{ d Y_W\,d Y_{b \bar{b}}}
 &=&\,\,\frac{1}{2\,\sigma_{eff}}  \frac{\nu^{SP}_1\Lb \mbox{LHC}\Rb \,\nu^{SP}_2\Lb \mbox{LHC}\Rb \,\nu'^{SP}_1\Lb \mbox{LHC}\Rb \,\nu'^{SP}_2\Lb \mbox{LHC}\Rb }{\nu^{SP}_1\Lb\mbox{Tevatron}\Rb \,\nu^{SP}_2\Lb\mbox{Tevatron}\Rb\,\nu'^{SP}_1\Lb\mbox{Tevatron}\Rb\,\nu'^{SP}_2\Lb\mbox{Tevatron}\Rb}
\notag \\
 &\times &
 \frac{ d
\sigma_{incl}\Lb \ln(s/M^2_{b , \bar{b}});Y_{b \bar{b}}; 0 \Rb}{d
Y_{b\bar{b}}}
    \,\,\frac{d \sigma_{incl}\Lb \ln(s/M^2_W);Y_{W};0\Rb}{ d Y_{W}}\,\label{EB1}\\
 &\approx & \frac{1}{16\,\sigma_{eff}}\,\frac{ d
\sigma_{incl}\Lb \ln(s/M^2_{b , \bar{b}});Y_{b \bar{b}}; 0 \Rb}{d
Y_{b\bar{b}}}
    \,\,\frac{d \sigma_{incl}\Lb \ln(s/M^2_W);Y_{W};0\Rb}{ d Y_{W}}\notag
\eea

The ratio of this background to the signal, can be written as
follows

\bea \frac{M^2_{b \bar{b}}\, \frac{ d^2\sigma \Lb p + p \to W + [b
\bar{b}] + X\Rb}{ d Y_W\,d Y_{b \bar{b}}}}{\frac{d^2 \sigma \Lb p +
p \to W +H + X\Rb}{d Y_W \,d Y_H}} &=& \frac{\frac{d
\sigma_{incl}\Lb p + p \to W + X\Rb}{ d Y_{W}}}{\frac{d^2
\sigma_{incl}\Lb p + p \to W + H + X \Rb}{ d Y_{W}\,d
Y_H}}\,\,\frac{1}{16\,\sigma_{eff}}\, \,M^2_{b \bar{b}}\frac{ d^2
\sigma_{incl}\Lb  p + p \to b \bar{b} + X\Rb}{d  M^2_{b
\bar{b}}\,\,d Y_{b\bar{b}}}\label{EB2} \eea

The first factor in \eq{EB2} is well known  (see \eq{WH2W}), and the
uncertainties in the values of the gluon structure function does not
affect it, to within an accuracy of 7\%. To evaluate the cross
section $M^2_{b \bar{b}}\frac{ d^2 \sigma_{incl}\Lb  p + p \to b
\bar{b} + X\Rb}{d  M^2_{b \bar{b}}\,\,d Y_{b\bar{b}}}$, we rewrite
this expression as follows

\beq \label{EB3} M^2_{b \bar{b}}\frac{ d^2 \sigma_{incl}\Lb p + p
\to b \bar{b} + X\Rb}{d  M^2_{b \bar{b}}\,\,d
Y_{b\bar{b}}}\,\,\,=\,\,\frac{\sigma\Lb M^2_{b
\bar{b}}\Rb}{\sigma\Lb M^2_H\Rb}\,\ \frac{ d \sigma \Lb p + p \to H
+ X\Rb}{d Y_H} \eeq

The values of  $ \sigma\Lb M^2_{b \bar{b}}\Rb$ and $\sigma\Lb
M^2_H\Rb$ are calculated in Appendix 2 and Appendix 3 and it is
equal to

 \beq \label{EB4} \frac{\sigma\Lb M^2_{b
\bar{b}}\Rb}{\sigma\Lb M^2_H\Rb}\,\,=\,\,\frac{4}{3}\,\frac{32
\as^2}{M^2_{b \bar{b}} \,A^2}\,\ln \Lb M^2_{b \bar{b}}/4m^2_b \Rb
\,\,\,\approx\,\,\,357 \eeq

 We believe that this ratio, which we
obtain in LO QCD, does not depend on the accuracy of our
calculation, and will be almost the same as in the NLO approximation
of perturbative QCD. All the uncertainties related to the
calculations in high order perturbative QCD in \eq{EB3}, are
absorbed in the value of the inclusive Higgs cross section.  For
this value, we take 3 \,$\times\,10^{-9}\,\mbox{mb}$ (for LO QCD),
and 5 \,$\times\,10^{-9}\,\mb$ (for NLO QCD) (see Refs.
\cite{NLOCAL}. Substituting this value and \eq{EB4} into \eq{EB3},
we obtain

\beq \label{EB5} M^2_{b \bar{b}}\frac{ d^2 \sigma_{incl}\Lb  p + p
\to b \bar{b} + X\Rb}{d M^2_{b \bar{b}}\,\,d
Y_{b\bar{b}}}\,\,\,=\,\,\,(1\,\div 2)\,\times\,\,10^{-6}\,\,\mb \eeq

 Using this
value, we estimate the background-to-signal ratio, as

\beq \label{EB6} \frac{M^2_{b \bar{b}}\, \frac{ d^3 \sigma \Lb p + p
\to W + [b \bar{b}] + X\Rb}{ d Y_W\,d Y_{b \bar{b}}}}{\frac{d \sigma
\Lb p + p \to W +H + X\Rb}{d Y_W \,d Y_H}} \,\,\,=\,\,\frac{1.24
\div 2.4}{16} \eeq

 Therefore, if \eq{FED7} is responsible for the two
parton shower production, the background in  the LHC kinematic
region will be small, and it can be neglected.

However, if the proton has some non-perturbative contribution,  and
differs from the colourless dipole of perturbative QCD, we can hope
that the non-enhanced diagram gives the main contribution, leading
to

 \beq \label{EB7} \frac{M^2_{b \bar{b}}\, \frac{ d^3 \sigma \Lb p
+ p \to W + [b \bar{b}] + X\Rb}{ d Y_W\,d Y_{b \bar{b}}}}{\frac{d
\sigma \Lb p + p \to W +H + X\Rb}{d Y_W \,d Y_H}} \,\,\,=\,\,1.24
\div 2.4 \eeq

The background of \eq{EB7}, should be considered . However,  using
the fact that the production from the two parton showers does not
depend on rapidity, while the signal leads to the sharp rapidity
distribution in the rapidity of the $b \bar{b}$ pair, the background
can be easily separated. We would like to draw your attention to the
fact that all our estimates, are  heavily based on the beautiful
measurement, of the so called double parton cross section,  by the
CDF collaboration.     We  would like to draw the attention of
experimentalists, to the fact that this kind of measurements in the
LHC region, will clarify the value of the contribution of the two
parton shower processes, and therefore, will lead to an improvement
of our knowledge of the non-traditional  QCD background.

\section{Summary}
This paper has two main results. The first, are the estimates of the
background to the associate W Higgs production. The second result,
is the detailed calculation of the contribution of the enhanced
diagrams to this background. As far as practical estimates are
concerned, \eq{EB6} and \eq{EB7}  give  the main  conclusions  of
this paper.  These results are encouraging, since the background
turns out to be no larger, than a factor 2 larger than the signal.
The principle difference between the signal and background, is the
dependence with respect to the rapidity difference $Y_W  - Y_{b
\bar{b}}$. The signal has a peak at $Y_W  - Y_{b \bar{b}} \,=\,0$,
while the background does not depend on this difference. Having this
in mind, we can easily separate this background for the signal.

The theoretical calculations, are the first where all kinematic
regions in the enhanced diagram have been considered. The most
interesting part, is related to the calculation of the contribution
of the overlapping singularities. Being interesting theoretically,
these singularities give a negligible contribution at the LHC
energies and below. We believe, that this is an important
observation for practical estimates of the processes, that are
responsible for physics in the saturation domain.

We would like to draw the attention of  experimentalists, to the
fact that the measurement of the so called double parton cross
section in the LHC range of energies, in the way that the CDF
collaboration did at the Tevatron, will clarify both the
contribution to the background of  such non-traditional QCD
processes as the two parton shower production, and  the importance
of the shadowing corrections for semihard processes.

We hope that our paper will be useful for planning experiments, in
the search for the Higgs boson.

\section*{Acknowledgments}
We want to thank Asher Gotsman,  Uri Maor and  Alex Prygarin  for
very useful discussions on the subject of this paper.

 This research was supported in part  by the Israel Science Foundation,
founded by the Israeli Academy of Science and Humanities, by a grant
from the Ministry of Science, Culture \& Sport, Israel \& the
Russian Foundation for Basic research of  the Russian Federation,
and by BSF grant \# 20004019. One of us (E.L.), wants to thank
Humboldt Foundation for financial support, that allowed him to start
this paper during his visit to DESY.

\renewcommand{\theequation}{A-1\arabic{equation}}
\setcounter{equation}{0}  
\begin{appendix}
\section{Appendix}

\numberwithin{equation}{subsection}
\subsection{The amplitude for the subprocess in
$GG\,\,\rightarrow\,\,W$}

 The aim of
this section of the appendix, is to derive the  expression for the
cross section $\sigma(M_{b \bar{b}})$, for the production of the $b
\bar{b}$ quark - anti quark pair, in the gluon fusion $ G G \to b
\bar{b}$. These cross sections, have been calculated (see
Refs.\cite{WXS,BXS,HXS}), but we reproduce the calculation here in
our technique using Sudakov variables, since using this method of
calculation,  it is easier to include the calculated cross sections
in the framework of our approach, based on the BFKL Pomeron.

 In \fig{hex}, one can see that in the Born diagram for this process,
the hexagon is on shell, as well as the protons between the
t-channel gluons. In other words, the dashed line through the center
of the diagram, intersects lines representing particles which are on
shell. The cross section which corresponds to this diagram, is equal
to

\beq \label{AXS} \frac{d\,\sigma^{BA}}{d Y_W} \,\,=\, \,\frac{1}{4
s}\,A\Lb \mbox{Mueller\, diagram of \fig{hex} for process p p W $
\to $ p p W}\Rb\,\,\delta \Lb\frac{1}{2}\ln\Lb \alpha_1/\beta_2\Rb -
y_W \Rb \eeq

 where $\alpha $ and $\beta$ are the Sudakov variables for
the vectors $\vec{q}_1$ and $\vec{q}_2$ ( see below). The upper line
of the diagram gives the factor

\beq
\Lb\,\frac{-ig_s}{q_1^2}\Rb\,\bar{u}\Lb\,p_1\,,\,s_1\,\Rb\,\gamma^{\mu}\,u\Lb\,\,p_1\,-q_1\,\,,\,s_1'\Rb\,\,\times\,\Lb\,\frac{-ig_s}{q_1^2}\Rb\,\bar{u}\Lb\,p_1-q_1\,,\,s_1'\,\Rb\,\gamma^{\rho}\,u\Lb\,\,p_1\,\,\,,\,s_1"\Rb\,\label{q}
\eeq

where $s_1$, $s_1'$ and $s_1"$ are the helicities of the incoming,
the intermediate, and the outgoing quarks respectively. Throughout
this calculation, the Eikonal approximation is used, which assumes
that the components of the momentum of the exchanged gauge particle,
(in this case gluons),  are small compared with the momentum of the
emitting quark. Hence, in the Eikonal approach, all components of
$\vec{q}_1$ are assumed to be small, such that \eq{q} may be
replaced by

\beq
\Lb\,\frac{4\pi\as}{q_1^4}\Rb\,\bar{u}\Lb\,p_1\,,\,s_1\,\Rb\,\gamma^{\mu}\,u\Lb\,\,p_1\,\,\,,\,s_1'\Rb\,\,\times\,\bar{u}\Lb\,p_1\,,\,s_1'\,\Rb\,\gamma^{\rho}\,u\Lb\,\,p_1\,\,\,,\,s_1"\Rb\,\label{qn}
\eeq

For spinors which are normalised such that
$u\Lb\,p_1\,,\,s_1'\Rb\,u^{\dag}\Lb\,p_1\,,\,s_1\Rb\,\,=\,\delta_{s_1,s_1'}$,
and using the following property of Dirac spinors from Gordon's
relation, namely

\beq
2\,\bar{u}\Lb\,p\,,\,s\Rb\,\gamma^{\mu}\,u\Lb\,p'\,,\,s'\Rb\,=\,\bar{u}\Lb\,p\,,\,s\Rb\,\Lb\,\Lb\,p+p'\Rb^{\mu}+i\sigma^{\mn}\Lb\,p-p'\Rb_{\nu}\Rb\,u\Lb\,p\,,\,s\Rb\,\label{gord}
\eeq

then plugging this into \eq{qn}, one has for the contribution of the
upper line of the diagram

\beq
\frac{\Lb\,4\pi\as\Rb\,}{q_1^4}\,p_1^{\mu}p_1^{\rho}\delta_{s_{1},s_1'}\,\delta_{s_{1}',s_1"}\,\,=\,\,
\frac{\Lb\,4\pi\as\Rb\,}{q_1^4}\,\frac{q^{\mu}_{1,\perp}}{\alpha_1}\,\frac{q^{\rho}_{1,\perp}}{\alpha_1}\,\delta_{s_{1},s_1'}\,\delta_{s_{1}',s_1"}
\label{q1} \eeq

where we use the gauge invariant condition, that
$q_1^{\mu}\,H_{\mu,\rho,\nu,\sigma}\,\,=\,\,0$, where $H$ is the
amplitude of the hexagon. Similarly, the lower line of the diagram
in the Eikonal approximation, gives the following contribution to
the amplitude

\beq
\frac{\Lb\,4\pi\as\Rb\,}{q_2^4}\,p_2^{\nu}p_2^{\sigma}\delta_{s_{2},s_2'}\,\delta_{s_{2}',s_2"}\,
\,=\,\,\frac{\Lb\,4\pi\as\Rb\,}{q_2^4}\,\,\frac{q^{\nu}_{2,\perp}}{\beta_2}\,\frac{q^{\sigma}_{1,\perp}}{\beta_2}\,\delta_{s_{2},s_2'}\delta_{s_{2}',s_2"}\label{q2}
\eeq

Hence, the expression for the amplitude of the diagram shown in
\fig{hex}, is given by the following expression, where it is assumed
that the on shell quarks (horizontal lines in the hexagon shown in
\fig{hex}), carry momenta which depend mostly on the longitudinal
direction, and all other quark lines in the hexagon, carry momenta
which depend mostly on the transverse direction,

\bea
A\Lb \fig{hex} \Rb\,&=&\,\frac{\Lb\,4\pi\as\Lb q^2\Rb\Rb^2}{q_1^4q_2^4\,\alpha_1^2\,\beta_2^2}\,\int\!\frac{d^4k}{\Lb 2\pi\Rb^4}\,\int\!\frac{d^4l}{\Lb2\pi\Rb^4\,}\,H\label{M} \\
&\,&\times\,\Lb 2\pi\Rb^2\delta\Lb\,-\,\beta_k\alpha_1\,s-\Lb\,\vec{k}-\vec{q}_1\Rb_{\bot}^2-m_u^2\Rb\,\delta\Lb\,\alpha_l\beta_2\,s-\Lb\,\vec{l}+\vec{q}_2\Rb_{\bot}^2-m_d^2\Rb\,\notag\\
\notag\\
\mbox{where}\,\,H\,\,&\equiv &\,q^{\mu}_{1,\bot}\,,q^{\rho}_{1,\bot}\,,q^{\nu}_{2,\bot}\,\,,q^{\sigma}_{2,\bot}\,H_{\mu\rho\nu\sigma} \notag \\
\,\,\,\,\,\,&\,=&\frac{f\Lb\,M_W\Rb}{D_1^2D_2^2}\,\varepsilon^{\lambda}\varepsilon^{\tau}\mbox{tr}\Lb\,\s{q}_2\s{k}\,\gamma_{\lambda}\Lb\,1-\gamma_5\Rb\,\s{l} \s{q}_1 \Lb\s{l}-\s{q}_1\Rb \s{q}_1\s{l}\gamma_{\tau}\Lb\,1-\gamma_5\Rb\,\s{k} \s{q}_2\Lb\s{k}+\s{q}_2\Rb\Rb\,\notag\\
&\,=&\frac{2f\Lb\,M_W\Rb}{D_1^2D_2^2}\,\,z\alpha_1\,\beta_2\,s \,\,q^2_1 \,q^2_2\varepsilon^{\lambda}\varepsilon^{\tau}\mbox{tr}\Lb\,\s{k}\gamma^{\bot}_{\lambda}\s{l}\s{l}\gamma^{\bot}_{\tau}\Lb\,1-\gamma_5\Rb\,\s{k}\Rb\,\notag\\
&\,=&\frac{8f\Lb\,M_W\Rb}{D_1^2D_2^2}\,\,\alpha_1\,\beta_2 s \,\,q^2_1 \,q^2_2k^2\,l^2 \notag\\
\label{AH}\\
\mbox{and
where}\,\,\sigma\Lb\,M_W\Rb\,&\,=\,&\,4\Lb\,4\pi\as\Lb M_W\Rb\Rb^2\,\,\sqrt{2}G_F\,M_W^2\Lb\frac{N_c^2-1}{2N_c}\Rb^2\frac{1}{4N_c}\notag\\
D_1\,&=&\,k^2-m_u^2\,\,=\,\alpha_k\beta_k\,s-k_{\bot}^2-m_u^2\notag\\
\,D_2\,&=&\,l^2-m_d^2\,\,\,\,=\,\alpha_l\beta_l\,s-l_{\bot}^2-m_d^2\label{AHN}\eea

where in the last step, $\gamma_5$ anticommutes past 4 gamma
matrices in the trace, to give a $\Lb\,1-\gamma_5\Rb^2
\,=\,2\Lb\,1-\gamma_5\Rb\,$ term. We use the fact that the part of
the trace on the RHS of \eq{M}, which contains a $\gamma_5$ matrix
can be removed, because evaluating the part of the trace containing
the $\gamma_5$ matrix gives a result which is antisymmetric in $\mu$
and $\rho$, which contracted with the symmetric term
$p_1^{\mu}p_1^{\rho}$, gives zero. $\varepsilon^{\lambda}$ and
$\varepsilon^{\lambda}$ are the polarisation vectors of the final
state W bosons in \fig{hex}. In \eq{M}, we use the fact that

\beq
\varepsilon^{\lambda}\varepsilon^{\tau}\,=\,g^{\lambda\tau}\label{pol}
\eeq

and using the integration measure

\beq \int\!\frac{d^4k}{\Lb 2\pi\Rb^4}\int\!\frac{d^4l}{\Lb
2\pi\Rb^4}\,=\,\frac{s}{2}\int\!d\alpha_k\,d\beta_k\frac{d^2k_{\bot}}{\Lb
2\pi\Rb^4}\frac{s}{2}\int\!d\alpha_ld\beta_l\frac{d^2l_{\bot}}{\Lb
2\pi\Rb^4}\label{intmes}\eeq

then \eq{M} simplifies to

\bea
A\Lb\,\fig{hex}\,\Rb\,&=&\,\frac{\Lb \,4\pi\as\Lb q^2\Rb\Rb^2}{q^4_1\,q^4_2\alpha_1^2\beta_2^2}\,\frac{s}{2}\int\!d\alpha_k\,d\beta_k\,\frac{d^2k_{\bot}}{\Lb 2\pi\Rb^3}\,\frac{s}{2}\int\!d\alpha_l\,d\beta_l\,\frac{d^2l_{\bot}}{\Lb 2\pi\Rb^3}\,\,\,H\notag\\
&\,&\times\,\delta\Lb\,-\,\beta_k\alpha_1\,s-\Lb\,\vec{k}-\vec{q}_1\Rb_{\bot}^2-m_u^2\Rb\,\delta\Lb\,\alpha_l\beta_2\,s-\Lb\,\vec{l}+\vec{q}_2\Rb_{\bot}^2-m_d^2\Rb\,\label{M2}
\eea

Using all the $\delta$ functions to evaluate the relevant
integrations, specifically taking the integral over $\alpha_1
\,>\,\alpha_l$ and over $\beta_2 \, >\,\beta_k$,  and using
\eq{AXS}, we obtain

 \bea\label{Ad} \frac{d\sigma^{BA}}{d Y_W}\,
&=&\,\, \,\,\frac{ \as^2(q^2)}{q^2_{1,\perp}\,\,q^2_{2,\perp}}
\,\sigma\Lb M_W\Rb \,\,\,\,\,\, \notag\\
\mbox{where}\,\,\,\,\,\,\,\,\,\,\,\,\,\,\,\,\,\,\,\,\,\,\,\,\,\,\,\,\,\,\,
\sigma\Lb M_W\Rb &=& \Lb\frac{N^2_c - 1}{2\,N_c}\Rb^2 \frac{1}{4\,
N_c}\frac{\as^2}{4
\,\pi^3}\,\sqrt{2}\,G_F\ln\Lb\frac{M_W^2}{4m_u^2}\Rb\ln\Lb\frac{M_W^2}{4m_d^2}\Rb\eea

It should be stressed, that the factor  $ \as^2(q^2)/\Lb
q^2_{1,\perp}\,\,q^2_{2,\perp}\Rb $, is included in the exchange of
the BFKL Pomeron, or in more general words, it should be absorbed in the product of the two gluon functions\\
 $x_1G(q^2_1,x_1)\,
x_2G(q^2_2,x_2)$ that give the flux, (luminosity) of the gluons from
both protons. The origin of the logs in \eq{Ad}, comes from the
integrations over $k_{\bot}$ and $l_{\bot}$. These integrations take
the following form

\beq \label{LOG} \int\,\,d^2 k_{\bot} k^2_{\bot} \Lb\mbox{ from the
tr(\dots)} \Rb\,\frac{1}{D^2_1}\,\,=\,\,\pi \,\,\ln (M^2_W/4 m^2_u)
\eeq

\renewcommand{\theequation}{A-2\arabic{equation}}
\numberwithin{equation}{subsection}
\subsection{The amplitude for subprocess
$GG\rightarrow\,\,b\bar{b}$}

In this appendix, we calculate the  amplitude for the subprocess
$GG\rightarrow\,\,b\bar{b}$, in the region of large values of $b
\bar{b}$ mass, ( denoted by $M_{b \bar{b}} \,\approx\,M_H$). For
such large values of the $b\bar{b}$ mass, we can restrict ourselves
by finding the contribution of the order of   $\ln \Lb  M^2_{b
\bar{b}}/4\,m^2_b \Rb$, which stems from the simple diagram of
\fig{bbin}. The cross section can be calculated using the following
expression

\bea \label{AXS1}
M^2_{b \bar{b}}\,\frac{d^2 \sigma^{BA}}{d M^2_{b \bar{b}}\,d Y_{b \bar{b}}}\,\,&=&\,\,\frac{1}{4\,s}\,\,A\Lb \mbox{Mueller amplitude (see \fig{bbin})  for the processes p + p + $ [b \bar{b}] \to$  p + p + $ [b \bar{b}]$}\Rb \notag \\
 &\times & \,\,M^2_{b \bar{b}}\,\delta \Lb \alpha_1\,\beta_2\,s -  M^2_{b \bar{b}}\Rb \,\delta \Lb \frac{1}{2}\,\ln \Lb \alpha_1/\beta_2\Rb \,-\,Y_{b \bar{b}}\Rb
\eea

The amplitude for this
 diagram  is given by the expression

\bea
A\Lb \fig{bbin} \Rb\,\,&=&\,\ikk\frac{\Lb-ig_s\Rb}{q_1^2}\bar{u}\Lb\,p_1\,,\,s_1\Rb\,\gamma^{\mu}u\Lb\,p_1'\,\,s_1'\Rb\,\frac{\Lb -ig_s\Rb}{q_2^2}\bar{u}\Lb\,p_2\,,\,s_2\Rb\,\gamma^{\nu}u\Lb\,p_2'\,\,s_2'\Rb\,\label{bb1}\\
&\times\,&\,\Lb\,\frac{\Lb-ig_s\Rb}{q_1^2}\bar{u}\Lb\,p_1\,,\,s_1\Rb\,\gamma^{\rho}u\Lb\,p_1'\,\,s_1'\Rb\,\frac{\Lb-ig_s
\Rb}{q_2^2}\bar{u}\Lb\,p_2\,,\,s_2\Rb\,\gamma^{\sigma}u\Lb\,p_2'\,\,s_2'\Rb\,\Rb^{\dag}S_{\mu\nu\rho\sigma}\notag
\eea

Using \eq{q1} and \eq{q2}, we find

\bea
A\Lb \fig{bbin} \Rb &=&  \frac{4 \pi  \as}{\alpha_1^2 \,\,q^4_1}\,  \frac{4 \pi  \as}{\beta^2_2\,\,q^4_2}\,\,  \ikk \,\,S \notag \\
\mbox{where}\,\,S\,&=&
\,\frac{\Lb4\pi\as\Lb\,M_{b\bar{b}}\Rb\Rb^2}{\Lb
k^2-m_b^2\Rb^2}q_{1,\bot}^{\mu}q^{\rho}_{1,\bot}q^{\nu}_{2,\bot}q^{\sigma}_{2,\bot}\tr\left\{\Lb\s{k}-\s{q}_1+m_b\Rb\gamma_{\rho}\Lb\s{k}+m_b\Rb\gamma_{\sigma}\Lb\s{k}+\s{q}_2+m_b\Rb\gamma_{\nu}\Lb\s{k}+m_b\Rb\gamma_{\mu}\right\}\notag\\
&\times\,&\Lb 2\pi\Rb^2\delta\Lb \Lb \vec{k}-\vec{q}_1\Rb^2-m_b
\Rb\,\,\delta\Lb \Lb \vec{k}+\vec{q}_2\Rb^2-m_b\Rb\,\Lb \frac{N^2_c
- 1}{ N_c} \Rb^2\,\frac{1}{4\,N_c} \label{S} \eea

In \eq{S},  we have used the Sudakov expansion

 \FIGURE[h]{\begin{minipage}{60mm}
\centerline{\epsfig{file=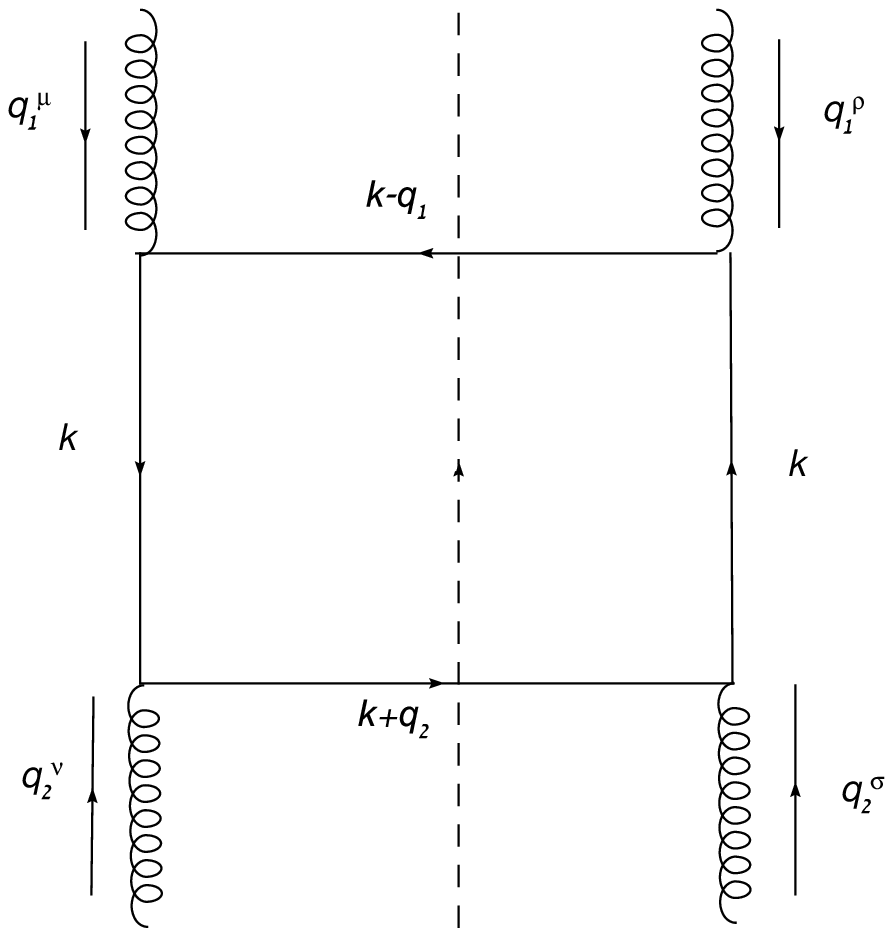,width=50mm}}
\end{minipage}
\caption{ The process $G + G   \to   b  + \bar{b}$.} \label{bbsq} }

\bea
\vec{k}\,&=&\,\alpha_k\vec{p}_1+\beta_k\vec{p}_2+\vec{k}_{\bot}\notag\\
\vec{q}_1\,&=&\,\alpha_1\vec{p}_1+\beta_1\vec{p}_2+\vec{q}_{1\bot}\,\notag \\
\vec{q}_2\,&=&\,\alpha_2\vec{p}_1+\beta_2\vec{p}_2+\vec{q}_{2\bot}\label{sudbb}
\eea

To calculate the trace in the expression for $S$ given in \eq{S}, we
recall  that at high energy, the horizontal on mass shell b quarks
in \fig{bbsq}, have momentum which depend mostly on the longitudinal
direction, and the vertical virtual b quark propagators, have
momentum which depend mostly on the transverse direction. This
observation reduces the RHS of \eq{S} to the following expression

\bea \label{TRACEB}
&&\tr\left\{\Lb\s{k}-\s{q}_1+m_b\Rb\s{q}_{1,\bot}\Lb\s{k}+m_b\Rb\,\s{q}_{2,\bot}\Lb\s{k}+\s{q}_2+m_b\Rb
\s{q}_{2,\bot}\Lb\s{k}+m_b\Rb\s{q}_{1,\bot}\right\}\,\,=\,\,4\,\alpha_1\,\beta_2\,s\,q^2_1\,q^2_2\Lb
\,k^2 + m^2_b\Rb\notag\\
~ \eea

In terms of the Sudakov parametres introduced in \eq{sudbb}, the
integration measure of the $k$ momentum space, becomes

\beq
\int\!\frac{d^4k}{\Lb2\pi\Rb^4}\,=\,\frac{s}{2}\int\!\frac{d^2k_{\bot}d\rho\,d\lambda}{\Lb2\pi\Rb^4}\label{intm}
\eeq

Using \eq{intm},  we integrate \eq{bb1} over all the $\alpha$-s and
$\beta$-s, taking into account the $\delta$-functions in \eq{AXS1}.
The result is given by the following expression

\bea M^2_{b \bar{b}}\,\frac{d^2 \sigma^{BA}}{d M^2_{b \bar{b}}\,d
Y_{b \bar{b}}}\,\,&=& \,\,\frac{
\as^2(q^2)}{q^2_{1,\perp}\,\,q^2_{2,\perp}}
\,\sigma\Lb M_{b \bar{b}}\Rb \,\,\,\,\,\, \notag\\
\mbox{where}\,\,\,\,\,\,\,\,\,\,\,\,\,\,\,\,\,\,\,\,\,\,\,\,\,\,\,\,\,\,\,
\sigma\Lb M_{b \bar{b}}\Rb &=& \Lb\frac{N^2_c - 1}{2 N_c}\Rb^2
\frac{1}{16\, N_c}\frac{\as^2}{8 \,\pi}\, \frac{1}{M^2_{b
\bar{b}}}\,\ln\Lb\frac{M_W^2}{4m_b^2}\Rb \label{Ab} \eea

where the log term has the same origin as in \eq{LOG}.

\subsection{The amplitude for the inclusive Higgs production via  subprocess
$GG\rightarrow\,\,H$}

The expression for the cross section for the inclusive Higgs
production process shown in \fig{HiggsInc}, is very similar to the
process of the $W$ production due to gluon fusion, and it is equal
to

\beq \label{AXSH} \frac{d\,\sigma^{BA}}{d Y_H} \,\,=\, \,\frac{1}{4
s}\,A\Lb \mbox{Mueller\, diagram of \fig{HiggsInc} for the process p
p H $ \to $ p p  H}\Rb\,\,\delta \Lb\frac{1}{2}\ln\Lb
\alpha_1/\beta_2\Rb - y_H \Rb \eeq

The amplitude for this process shown in \fig{HiggsInc}, is given by

\bea A\Lb \fig{HiggsInc}\Rb\,&=&
\,A^2\Pi_{\mn}\Pi_{\rho\sigma}\frac{\Lb 4\pi\as\Lb
q^2\Rb\Rb^2}{q_1^4q_2^4}  \label{HI}\\
&\times \,&\,\sum_{\footnotesize{\mbox{spins}}}\left\{\bar{u}\Lb
p_1'\Rb\gamma^{\mu}u\Lb p_1\Rb\,\bar{u}\Lb p_2'\Rb\gamma^{\nu}u\Lb
p_2\Rb\right\}\left\{\bar{u}\Lb p_1'\Rb\gamma^{\rho}u\Lb
p_1\Rb\,\bar{u}\Lb p_2'\Rb\gamma^{\sigma}u\Lb
p_2\Rb\right\}^{\dag}\notag\eea

 \beq
\mbox{where}\,\,\,\,\,\,\,\,\,A^2\,=\,\frac{4\,\sqrt{2}G_F\as^2\Lb
M_H\Rb}{9}\Lb
N_C^2-1\Rb\,\,\,\,\,\,\,\,\,\mbox{and}\,\,\,\,\,\,\,\,\Pi^{\mn}\,=\,q_1^{\nu}q_2^{\mu}-\frac{M_H^2}{2}g^{\mn}\label{Ag}\eeq

 \FIGURE[h]{\begin{minipage}{65mm}
\centerline{\epsfig{file=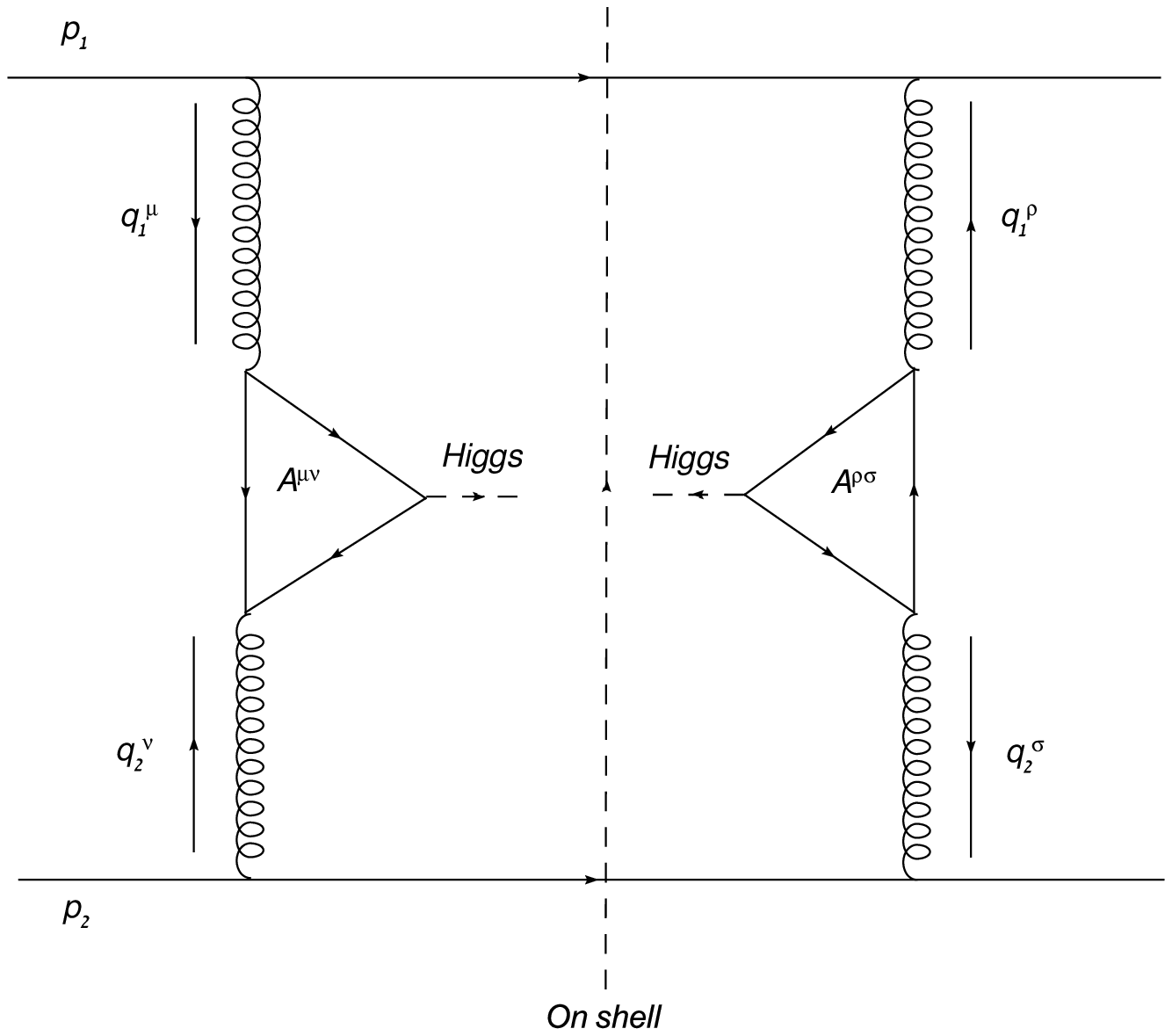,width=60mm}}
\end{minipage}
\caption{The process of inclusive Higgs production due to gluon
fusion: $G + G \to Higgs$.} \label{HiggsInc} }

Using \eq{q1} and \eq{q2}, as well as the form of the gauge
invariant tensor $\Pi^{\mn}$, we readily obtain the following
expression after integration over the $\alpha$-s and $\beta$-s

\bea \frac{d\,\sigma^{BA}}{d Y_H} \,\,&=&\,\,\frac{
\as^2(q^2)}{q^2_{1,\perp}\,\,q^2_{2,\perp}}
\,\sigma\Lb M_H\Rb \,\,\,\,\,\, \notag\\
\mbox{where}\,\,\,\,\,\,\,\,\,\,\,\,\,\,\,\,\,\,\,\,\,\,\,\,\,\,\,\,\,\,\,
\sigma\Lb M_{H}\Rb &=& \Lb\frac{N^2_c - 1}{4 N^2_c}\Rb
\frac{A^2}{256 \pi} \label{Ah} \eea

Since the  Pomeron exchange, or more specifically the gluon
structure functions enter in the same way for the case of the $b
\bar{b}$ pair production, as in the case for Higgs production due to
gluon fusion, we can calculate the ratio

\beq \frac{  M^2_{b \bar{b}}\frac{d^2 \sigma\Lb  p +p \to [ b
\bar{b}] \,+\,X \Rb}{d M^2_{b \bar{b}}\, d Y_{b \bar{b}}}}{
\frac{\sigma\Lb p + p \to H + X\Rb}{d Y_H}}\,\,=\,\,\frac{\sigma\Lb
M_{b \bar{b}}\Rb}{\sigma\Lb M_H\Rb} \eeq

 Using \eq{Ab} and \eq{Ad},
we have

 \beq \label{RATIO} M^2_{b \bar{b}}\frac{d^2 \sigma\Lb  p +p
\to [ b \bar{b}] \,+\,X \Rb}{d M^2_{b \bar{b}}\, d Y_{b \bar{b}}}
\,\,=\,\,\frac{4}{3}\,\frac{32 \as^2}{M^2_{b \bar{b}} \,A^2}\,\ln
\Lb M^2_{b \bar{b}}/4m^2_b \Rb\,\times\, \frac{d\sigma\Lb p + p \to
H + X\Rb}{d Y_H} \eeq

 where the factor of  $4/3$ is a colour factor.

\end{appendix}


\begin{thebibliography}{99}
\bibitem{DG}
V.~A.~Khoze, A.~D.~Martin and M.~G.~Ryskin,
  arXiv:0705.2314 [hep-ph] and references therein.

\bibitem{TXS}
 P.~V.~Landshoff,
  ``The total cross section at the LHC,''
  arXiv:0709.0395 [hep-ph], arXiv:hep-ph/0509240;\,\,\,
   A.~Achilli, R.~Hegde, R.~M.~Godbole, A.~Grau, G.~Pancheri and Y.~Srivastava,
  ``Total cross-section and rapidity gap survival probability at the LHC
  through an eikonal with soft gluon resummation,''
  arXiv:0708.3626 [hep-ph];\,\,\,\, K.~Igi and M.~Ishida,
  ``Predictions of p p, anti-p p total cross section and rho ratio at LHC and
  cosmic-ray energies based on duality,''
  arXiv:hep-ph/0510129, Phys.\ Lett.\  B {\bf 622} (2005) 286
  [arXiv:hep-ph/0505058];\,\,\,
G.~Matthiae,
  ``Total cross-section and luminosity,''
  Eur.\ Phys.\ J.\  C {\bf 4S1} (2002) 13.


\bibitem{DPXS}
  F.~Abe {\it et al.}  [CDF Collaboration],
  Phys.\ Rev.\  D {\bf 56}, 3811 (1997);\,\,
  Phys.\ Rev.\ Lett.\  {\bf 79}, 584 (1997).




\bibitem{FT}
 J.C. Collins, D.E. Soper and G. Sterman: Nucl. Phys. {\bf B308}, 833 (1988).
 \bibitem{KTF}
S.~Catani, M.~Ciafaloni and F.~Hautmann,
Nucl.\ Phys.\  {\bf B366}, 135 (1991).
E.~M.~Levin, M.~G.~Ryskin, Y.~M.~Shabelski and A.~G.~Shuvaev,
Sov.\ J.\ Nucl.\ Phys.\  {\bf 53}, 657 (1991)
[Yad.\ Fiz.\  {\bf 53}, 1059 (1991)];\,\,\,
J.~C.~Collins and R.~
K.~Ellis,
Nucl.\ Phys.\  {\bf B360}, 3 (1991).

\bibitem{AGK}
V.~A.~Abramovsky, V.~N.~Gribov and O.~V.~Kancheli,
  { \it Yad.\ Fiz.}\, {\bf 18}, 595 (1973)
  [{\it Sov.\ J.\ Nucl.\ Phys.}\, {\bf 18}, 308 (1974)].





\bibitem{BFKL}
 E. A. Kuraev, L. N. Lipatov, and F. S. Fadin, {\it  Sov. Phys. 
JETP}
                {\bf 45}, 199 (1977); \,\,\,
Ya. Ya. Balitsky and L. N. Lipatov,
               {\it   Sov. J. Nucl. Phys.}\, {\bf 28}, 22 (1978).


\bibitem{LMP}
 E.~Levin, J.~Miller and A.~Prygarin,
  arXiv:0706.2944 [hep-ph].

\bibitem{MUIN}
A.~H.~Mueller,
Phys.\ Rev.\  {\bf D 2}, 2963 (1970); Phys.\ Rev.\ {\bf D 4}, 150  (1971).
\bibitem{INXS}
E.~M.~Levin and M.~G.~Ryskin,
Phys.\ Rept.\  {\bf 189} (1990) 267;\,\,\,
E.~Laenen and E.~Levin,
{\it Nucl.\ Phys.}\,  {\bf B451} (1995) 207;
{\it Ann.\ Rev.\ Nucl.\ Part.\ Sci.}\,  {\bf 44} (1994) 199.
[arXiv:hep-ph/9503381];\,\,\,
M.Gyulassy and L. McLerran, {\it Phys. Rev.} {\bf C56} (1997) 2219;\,\,\,
Y.~V.~Kovchegov and K.~Tuchin,
Phys.\ Rev.\ D {\bf 65} (2002) 074026
[arXiv:hep-ph/0111362];


\bibitem{DGLAP}
 V. N. Gribov and L. N. Lipatov,  Sov. J. Nucl. Phys {\bf 15} (1972)
                438;\\
 G. Altarelli and G. Parisi, Nucl. Phys. {\bf B 126} (1977) 298; \\
Yu. l. Dokshitser, Sov. Phys. JETP {\bf 46}  (1977) 641.

\bibitem{GLR}
L. V. Gribov, E. M. Levin and M. G. Ryskin, {\it Phys. Rep.}\,
{\bf 100}, 1 (1983);\,\,\,
A. H. Mueller and J. Qiu,  {\it Nucl. Phys.},427 {\bf B 268}
(1986) ;\,\,\,
L. McLerran and R. Venugopalan, {\it  Phys. Rev.}  {\bf D 49},2233,
3352  (1994); {\bf D 50},2225 (1994); {\bf D 53},458 (1996); {\bf
D 59},09400
(1999).
\bibitem{BART}
J.~Bartels, M.~Braun and G.~P.~Vacca,
 { \it Eur.\ Phys.\ J.}  {\bf C40}, 419 (2005)
  [arXiv:hep-ph/0412218]\,;\,\,\,
J.~Bartels and C.~Ewerz,
{\it JHEP} {\bf 9909}, 026 (1999)
[arXiv:hep-ph/9908454]\,;\,\,\,
J.~Bartels and M.~Wusthoff,
{\it Z.\ Phys.} {\bf C66}, 157 (1995)\,;\,\,\,
\,\,\,\,A.~H.~Mueller and B.~Patel,
{\it Nucl.\ Phys.}  {\bf B425}, 471 (1994)
[arXiv:hep-ph/9403256];\,\,\,
J.~Bartels,
Z.\ Phys.\  {\bf C60}, 471 (1993).



\bibitem{BRN}
M.~A.~Braun,
{\it   Phys.\ Lett.}\,  {\bf B632} (2006) 297
  [arXiv:hep-ph/0512057];\,\,
arXiv:hep-ph/0504002\,;
{ \it Eur.\ Phys.\ J.}  {\bf C16}, 337 (2000)
[arXiv:hep-ph/0001268];\,\,\,
  Phys.\ Lett.\ B {\bf 483} (2000) 115
  [arXiv:hep-ph/0003004];\,\,
  Eur.\ Phys.\ J.\ C {\bf 33} (2004) 113
  [arXiv:hep-ph/0309293];\,\,\,
{\it Eur.\ Phys.\ J.}  {\bf C6}, 321 (1999)
[arXiv:hep-ph/9706373];\,\,\,
M.~A.~Braun and G.~P.~Vacca,
{\it Eur.\ Phys.\ J.}  {\bf C6}, 147 (1999)
[arXiv:hep-ph/97114].

\bibitem{BBXS}
 P.~Nason {\it et al.},
  ``Bottom production,'"
  arXiv:hep-ph/0003142;\,\,
  R.~Bonciani, S.~Catani, M.~L.~Mangano and P.~Nason,
  Nucl.\ Phys.\  B {\bf 529} (1998) 424
  [arXiv:hep-ph/9801375];\,\,
  S.~Frixione, M.~L.~Mangano, P.~Nason and G.~Ridolfi,
  Adv.\ Ser.\ Direct.\ High Energy Phys.\  {\bf 15} (1998) 609
  [arXiv:hep-ph/9702287],
  Nucl.\ Phys.\  B {\bf 431} (1994) 453.


\bibitem{NLOCAL}
S.~ Marsini,~R.~D.~ Ball, ~V.~ Del Ducca, ~S. ~Forte and ~A. ~Vicini,{\it `` Higgs production via gluon-gluon fusion with finite top mass beyond next-to-leading order"}, arXiv: 0801.2544[hep-ph] and reference therein;, {\it `` Inclusive Higgs boson production at hadron colliders at next-to-leading order"}, arXiv: hep-ph/0205152;\,\,\,
W.-M. Yao et al. (Particle Data Group), J. Phys. G 33, 1 (2006) and 2007 partial update for the 2008 edition.


\bibitem{WXS}
R.~Hamberg and W.~L.~van Neerven,
  Nucl.\ Phys.\  B {\bf 379} (1992) 143;\,\, T.~Matsuura, R.~Hamberg and W.~L.~van Neerven,
  Nucl.\ Phys.\ Proc.\ Suppl.\  {\bf 23B} (1991) 3;\,\,\, 
  Nucl.\ Phys.\  B {\bf 345} (1990) 331;\,\,\,
C.~Anastasiou, L.~J.~Dixon, K.~Melnikov and F.~Petriello,
  Phys.\ Rev.\  D {\bf 69} (2004) 094008
  [arXiv:hep-ph/0312266];\,\,\,  K.~Melnikov and F.~Petriello,
  Phys.\ Rev.\ Lett.\  {\bf 96} (2006) 231803
  [arXiv:hep-ph/0603182] asnd references therein.


\bibitem{BXS}
~P. ~Nason,~S. ~ Dawson and ~R.~K.~Ellis, Nucl.Phys. {\bf B 303}, 607 (1988), Nucl.Phys. {\bf B 327},49 (1989);
\,\,\,~M.~L.~Mangano, ~P.~Nason, and ~G.~Ridolfi, Nucl.Phys. {\bf B 373}, 295 (1992).

\bibitem{HXS}
J.~R. ~Ellis,~M,~K.~Gaillard and  ~D.~V. ~Nanopoulos, Nucl. Phys. {\bf B 106}, 292 (1976);\,\,\,
~A.~I.~Vainstein, ~M.~B.~Voloshin ans ~V.~I.~Zakharov, Sov. J. Nucl. Phys. {\bf 30}, 431 (1979);\,\,\,
The Review of Particle Physics
W.-M. Yao et al., Journal of Physics, G 33, 1 (2006)
and 2007 partial update for 2008.



\end{thebibliography}
\end{document}